%% file: Machine-learning_Growth_at_Risk.tex
\documentclass[12pt]{article}
\usepackage[english]{babel}
\usepackage[letterpaper,top=2cm,bottom=2cm,left=3cm,right=3cm,marginparwidth=1.75cm]{geometry}
\usepackage{amsmath, amsfonts, amsthm, bm}
\usepackage{graphicx}
\usepackage{enumerate}
\usepackage{setspace}
\usepackage{natbib}
\usepackage{float}
\usepackage{subfig}
\usepackage{multirow}
\usepackage{algorithm}
\usepackage{algpseudocode}
\usepackage{caption}
\usepackage{longtable, booktabs}

\newtheorem*{note*}{\protect\notename}
\providecommand{\notename}{Note}
\usepackage{environ} 
\usepackage{etoolbox} 

\input{tcilatex}
\begin{document}

\title{Machine-learning Growth at Risk\thanks{The views expressed herein are those of the authors and should not be
attributed to the IMF, its Executive Board, or its management. }\\
}
\date{May 2025}

\author{ \small{Tobias Adrian} \thanks{Monetary and Capital Markets Department, International Monetary Fund, tadrian@imf.org.} \and
        \small Hongqi Chen\thanks{College of Finance and Statistics, Hunan University, hongqichen@hnu.edu.cn.}  \and 
        \small Max-Sebastian Dovì\thanks{African Department, International Monetary Fund, mdovi@imf.org.} \and 
        \small Ji Hyung Lee\thanks{{Department of Economics, University of Illinois Urbana-Champaign, jihyung@illinois.edu.}}
        }



\maketitle

%

\begin{abstract}

\noindent We analyse growth vulnerabilities in the US using quantile partial correlation regression, a selection-based machine-learning method that achieves model selection consistency under time series. We find that downside risk is primarily driven by financial, labour-market, and housing variables, with their importance changing over time. Decomposing downside risk into its individual components, we construct sector-specific indices that predict it, while controlling for information from other sectors, thereby isolating the downside risks emanating from each sector.

\end{abstract}
\pagebreak

\input{sections/introduction_monthly}

\input{sections/methodology_monthly}

\input{sections/simulation}

\input{sections/empirics_monthly}

\input{sections/conclusion_monthly}

\pagebreak

\bibliographystyle{chicago}
\bibliography{AtRisk}

\pagebreak

\renewcommand{\thesection}{\Alph{section}}
\numberwithin{equation}{section}
\setcounter{section}{0}

\section{Appendix}

\subsection{Description of acronyms in Fred-MD}

{\scriptsize
\begin{longtable}{p{0.2\textwidth} p{0.5\textwidth} p{0.2\textwidth}}
\caption{Description of acronyms in FRED-MD}\\
\label{tab:monthly_dictionary}\\
\toprule
Acronym & Description & Group \\
\midrule
\endfirsthead
\multicolumn{3}{c}{{\bfseries \tablename\ \thetable{} -- continued from previous page}} \\
Mnemonic & Description & Group \\
\midrule
\endhead
\midrule \multicolumn{3}{r}{{Continued on next page}} \\
\endfoot
\bottomrule
\endlastfoot
RPI & Real Personal Income & 1 \\
W875RX1 & Real personal income ex transfer receipts & 1 \\
DPCERA3M086SBEA & Real personal consumption expenditures & 4 \\
CMRMTSPLx & Real Manu.  and Trade Industries Sales & 4 \\
RETAILx & Retail and Food Services Sales & 4 \\
INDPRO & IP Index & 1 \\
IPFPNSS & IP: Final Products and Nonindustrial Supplies & 1 \\
IPFINAL & IP: Final Products (Market Group) & 1 \\
IPCONGD & IP: Consumer Goods & 1 \\
IPDCONGD & IP: Durable Consumer Goods & 1 \\
IPNCONGD & IP: Nondurable Consumer Goods & 1 \\
IPBUSEQ & IP: Business Equipment & 1 \\
IPMAT & IP: Materials & 1 \\
IPDMAT & IP: Durable Materials & 1 \\
IPNMAT & IP: Nondurable Materials & 1 \\
IPMANSICS & IP: Manufacturing (SIC) & 1 \\
IPB51222s & IP: Residential Utilities & 1 \\
IPFUELS & IP: Fuels & 1 \\
CUMFNS & Capacity Utilization:  Manufacturing & 1 \\
HWI & Help-Wanted Index for United States & 2 \\
HWIURATIO & Ratio of Help Wanted/No.  Unemployed & 2 \\
CLF16OV & Civilian Labor Force & 2 \\
CE16OV & Civilian Employment & 2 \\
UNRATE & Civilian Unemployment Rate & 2 \\
UEMPMEAN & Average Duration of Unemployment (Weeks) & 2 \\
UEMPLT5 & Civilians Unemployed - Less Than 5 Weeks & 2 \\
UEMP5TO14 & Civilians Unemployed for 5-14 Weeks & 2 \\
UEMP15OV & Civilians Unemployed - 15 Weeks \& Over & 2 \\
UEMP15T26 & Civilians Unemployed for 15-26 Weeks & 2 \\
UEMP27OV & Civilians Unemployed for 27 Weeks and Over & 2 \\
CLAIMSx & Initial Claims & 2 \\
PAYEMS & All Employees:  Total nonfarm & 2 \\
USGOOD & All Employees:  Goods-Producing Industries & 2 \\
CES1021000001 & All Employees:  Mining and Logging:  Mining & 2 \\
USCONS & All Employees:  Construction & 2 \\
MANEMP & All Employees:  Manufacturing & 2 \\
DMANEMP & All Employees:  Durable goods & 2 \\
NDMANEMP & All Employees:  Nondurable goods & 2 \\
SRVPRD & All Employees:  Service-Providing Industries & 2 \\
USTPU & All Employees:  Trade, Transportation \& Utilities & 2 \\
USWTRADE & All Employees:  Wholesale Trade & 2 \\
USTRADE & All Employees:  Retail Trade & 2 \\
USFIRE & All Employees:  Financial Activities & 2 \\
USGOVT & All Employees:  Government & 2 \\
CES0600000007 & Avg Weekly Hours :  Goods-Producing & 2 \\
AWOTMAN & Avg Weekly Overtime Hours :  Manufacturing & 2 \\
AWHMAN & Avg Weekly Hours :  Manufacturing & 2 \\
HOUST & Housing Starts:  Total New Privately Owned & 3 \\
HOUSTNE & Housing Starts, Northeast & 3 \\
HOUSTMW & Housing Starts, Midwest & 3 \\
HOUSTS & Housing Starts, South & 3 \\
HOUSTW & Housing Starts, West & 3 \\
PERMIT & New Private Housing Permits (SAAR) & 3 \\
PERMITNE & New Private Housing Permits, Northeast (SAAR) & 3 \\
PERMITMW & New Private Housing Permits, Midwest (SAAR) & 3 \\
PERMITS & New Private Housing Permits, South (SAAR) & 3 \\
PERMITW & New Private Housing Permits, West (SAAR) & 3 \\
ACOGNO & New Orders for Consumer Goods & 4 \\
AMDMNOx & New Orders for Durable Goods & 4 \\
ANDENOx & New Orders for Nondefense Capital Goods & 4 \\
AMDMUOx & Unfilled Orders for Durable Goods & 4 \\
BUSINVx & Total Business Inventories & 4 \\
ISRATIOx & Total Business:  Inventories to Sales Ratio & 4 \\
M1SL & M1 Money Stock & 5 \\
M2SL & M2 Money Stock & 5 \\
M2REAL & Real M2 Money Stock & 5 \\
BOGMBASE & Monetary Base & 5 \\
TOTRESNS & Total Reserves of Depository Institutions & 5 \\
NONBORRES & Reserves Of Depository Institutions & 5 \\
BUSLOANS & Commercial and Industrial Loans & 5 \\
REALLN & Real Estate Loans at All Commercial Banks & 5 \\
NONREVSL & Total Nonrevolving Credit & 5 \\
CONSPI & Nonrevolving consumer credit to Personal Income & 5 \\
S.P.500 & S\&P’s Common Stock Price Index: Composite & 8 \\
S.P.div.yield & S\&P’s Composite Common Stock: Dividend Yield & 8 \\
S.P.PE.ratio & S\&P’s Composite Common Stock: Price-Earnings Ratio & 8 \\
FEDFUNDS & Effective Federal Funds Rate & 6 \\
CP3Mx & 3-Month AA Financial Commercial Paper Rate & 6 \\
TB3MS & 3-Month Treasury Bill: & 6 \\
TB6MS & 6-Month Treasury Bill: & 6 \\
GS1 & 1-Year Treasury Rate & 6 \\
GS5 & 5-Year Treasury Rate & 6 \\
GS10 & 10-Year Treasury Rate & 6 \\
AAA & Moody’s Seasoned Aaa Corporate Bond Yield & 6 \\
BAA & Moody’s Seasoned Baa Corporate Bond Yield & 6 \\
COMPAPFFx & 3-Month Commercial Paper Minus FEDFUNDS & 6 \\
TB3SMFFM & 3-Month Treasury C Minus FEDFUNDS & 6 \\
TB6SMFFM & 6-Month Treasury C Minus FEDFUNDS & 6 \\
T1YFFM & 1-Year Treasury C Minus FEDFUNDS & 6 \\
T5YFFM & 5-Year Treasury C Minus FEDFUNDS & 6 \\
T10YFFM & 10-Year Treasury C Minus FEDFUNDS & 6 \\
AAAFFM & Moody’s Aaa Corporate Bond Minus FEDFUNDS & 6 \\
BAAFFM & Moody's Baa Corporate Bond Minus FEDFUNDS & 6 \\
TWEXAFEGSMTHx & Trade Weighted U.S. Dollar Index & 6 \\
EXSZUSx & Switzerland / U.S. Foreign Exchange Rate & 6 \\
EXJPUSx & Japan / U.S. Foreign Exchange Rate & 6 \\
EXUSUKx & U.S. / U.K. Foreign Exchange Rate & 6 \\
EXCAUSx & Canada / U.S. Foreign Exchange Rate & 6 \\
WPSFD49207 & PPI: Finished Goods & 7 \\
WPSFD49502 & PPI: Finished Consumer Goods & 7 \\
WPSID61 & PPI: Intermediate Materials & 7 \\
WPSID62 & PPI: Crude Materials & 7 \\
OILPRICEx & Crude Oil, spliced WTI and Cushing & 7 \\
PPICMM & PPI: Metals and metal products: & 7 \\
CPIAUCSL & CPI : All Items & 7 \\
CPIAPPSL & CPI : Apparel & 7 \\
CPITRNSL & CPI : Transportation & 7 \\
CPIMEDSL & CPI : Medical Care & 7 \\
CUSR0000SAC & CPI : Commodities & 7 \\
CUSR0000SAD & CPI : Durables & 7 \\
CUSR0000SAS & CPI : Services & 7 \\
CPIULFSL & CPI : All Items Less Food & 7 \\
CUSR0000SA0L2 & CPI : All items less shelter & 7 \\
CUSR0000SA0L5 & CPI : All items less medical care & 7 \\
PCEPI & Personal Cons.  Expend.:  Chain Index & 7 \\
DDURRG3M086SBEA & Personal Cons.  Exp:  Durable goods & 7 \\
DNDGRG3M086SBEA & Personal Cons.  Exp:  Nondurable goods & 7 \\
DSERRG3M086SBEA & Personal Cons.  Exp:  Services & 7 \\
CES0600000008 & Avg Hourly Earnings :  Goods-Producing & 2 \\
CES2000000008 & Avg Hourly Earnings :  Construction & 2 \\
CES3000000008 & Avg Hourly Earnings :  Manufacturing & 2 \\
UMCSENTx & Consumer Sentiment Index & 4 \\
DTCOLNVHFNM & Consumer Motor Vehicle Loans Outstanding & 5 \\
DTCTHFNM & Total Consumer Loans and Leases Outstanding & 5 \\
INVEST & Securities in Bank Credit at All Commercial Banks & 5 \\
VIXCLSx & VIX & 8 \\
INDPRO.1 & INDPRO.1 & 9 \\
A2P2 & 1-mo. Nonfinancial commercial paper A2P2/AA credit spread & 10 \\
ABCP & 1-mo. Asset-backed/Financial commercial paper spread & 11 \\
ABSI & Nonmortgage ABS Issuance (Relative to 12-mo. MA) & 12 \\
ABSSPREAD & BofAML Home Equity ABS/MBS yield spread & 11 \\
BAA2 & Moody's Baa corporate bond/10-yr Treasury yield spread & 10 \\
BDG & Broker-dealer Debit Balances in Margin Accounts & 12 \\
BONDGR & New US Corporate Debt Issuance (Relative to 12-mo. MA) & 12 \\
CARSPREAD & UM Household Survey: Auto Credit Conditions Good/Bad spread & 10 \\
CBCAR & Commercial Bank 48-mo. New Car Loan/2-yr Treasury yield spread & 10 \\
CBILL & 3-mo. Financial commercial paper/Treasury bill spread & 11 \\
CBPER & Commercial Bank 24-mo. Personal Loan/2-yr Treasury yield spread & 10 \\
CCDQ & S\&P US Bankcard Credit Card: 3-mo. Delinquency Rate & 10 \\
CCG & Consumer Credit Outstanding & 10 \\
CCINC & S\&P US Bankcard Credit Card: Excess Rate Spread & 10 \\
CG & Commercial Paper Outstanding & 11 \\
CILARGE & FRB Senior Loan Officer Survey: Tightening Standards on Large C\&I Loans & 10 \\
CISMALL & FRB Senior Loan Officer Survey: Tightening Standards on Small C\&I Loans & 10 \\
CITA & Commercial Bank C\&I Loans/Total Assets & 12 \\
CMBS & BofAML 3-5 yr AAA CMBS OAS spread & 11 \\
CMBSI & CMBS Issuance (Relative to 12-mo. MA) & 12 \\
COMMODLIQ & COMEX Gold/NYMEX WTI Futures Market Depth & 12 \\
CONTA & Commercial Bank Consumer Loans/Total Assets & 12 \\
CPH & FRB Commercial Property Price Index & 12 \\
CPR & Counterparty Risk Index (formerly maintained by Credit Derivatives Research) & 11 \\
CRE & FRB Senior Loan Officer Survey: Tightening Standards on CRE Loans & 10 \\
CRG & S\&P US Bankcard Credit Card: Receivables Outstanding & 10 \\
CTABS & ICE BofAML ABS/5-yr Treasury yield spread & 11 \\
CTERM & 3-mo./1-wk AA Financial commercial paper spread & 11 \\
CTF & ICE BofAML Financial/Corporate Credit bond spread & 11 \\
CTMBS & ICE BofAML Mortgage Master MBS/10-year Treasury yield spread & 11 \\
CWILL & FRB Senior Loan Officer Survey: Willingness to Lend to Consumers & 10 \\
D10 & 10-yr Constant Maturity Treasury yield & 12 \\
D2 & 2-yr Constant Maturity Treasury yield & 12 \\
DCOMM & Commercial Bank Total Unused C\&I Loan Commitments/Total Assets & 12 \\
DNET & Net Notional Value of Credit Derivatives & 12 \\
DURSPREAD & UM Household Survey: Durable Goods Credit Conditions Good/Bad spread & 10 \\
EQUITYLIQ & CME E-mini S\&P Futures Market Depth & 12 \\
FAILS & Treasury Repo Delivery Fails Rate & 11 \\
FAILSA & Agency Repo Delivery Failures Rate & 11 \\
FAILSC & Corporate Securities Repo Delivery Failures Rate & 11 \\
FAILSMBS & Agency MBS Repo Delivery Failures Rate & 11 \\
FC & Total Assets of Finance Companies/GDP & 12 \\
FCORP & Total Assets of Funding Corporations/GDP & 12 \\
FFR & Federal Funds Rate & 12 \\
FG & Finance Company Owned \& Managed Receivables & 10 \\
FINS & S\&P 500 Financials/S\&P 500 Price Index (Relative to 2-yr MA) & 12 \\
GSE & Total Agency and GSE Assets/GDP & 12 \\
GVL & FDIC Volatile Bank Liabilities & 11 \\
HDQBC & NY Fed Consumer Credit Panel: Loan Delinquency Status: Non-current (Percent of Total Balance) & 10 \\
HDQBNP & NY Fed Consumer Credit Panel: New Delinquent Loan Balances (Percent of Current Balance) & 10 \\
HDQBNRP & NY Fed Consumer Credit Panel: New Seriously Delinquent Loan Balances (Percent of Current Balance) & 10 \\
HH & Household debt outstanding/PCE Durables and Residential Investment & 12 \\
HOUSSPREAD & UM Household Survey: Mortgage Credit Conditions Good/Bad spread & 10 \\
HY & BofAML High Yield/Moody's Baa corporate bond yield spread & 10 \\
INS & Total Assets of Insurance Companies/GDP & 12 \\
ITA & Fed funds and Reverse Repurchase Agreements/Total Assets of Commercial Banks & 12 \\
JINC & 30-yr Jumbo/Conforming fixed rate mortgage spread & 10 \\
LHY & Markit High Yield (HY) 5-yr Senior CDS Index & 10 \\
LIBID & 3-mo. Interbank Deposit Spread (OBFR/LIBID-Treasury) & 11 \\
LIG & Markit Investment Grade (IG) 5-yr Senior CDS Index & 10 \\
LPH & CoreLogic National House Price Index & 12 \\
MBONDGR & New State \& Local Government Debt Issues (Relative to 12-mo.h MA) & 12 \\
MBSI & Total MBS Issuance (Relative to 12-mo. MA) & 12 \\
MCAP & S\&P 500, NASDAQ, and NYSE Market Capitalization/GDP & 12 \\
MDQ & MBA Serious Delinquencies & 10 \\
MG & Money Stock: MZM & 10 \\
MINC & 30-yr Conforming Mortgage/10-yr Treasury yield spread & 10 \\
MLIQ10 & On-the-run vs. Off-the-run 10-yr Treasury liquidity premium & 11 \\
MMF & FMMFPIAN@USECON & 11 \\
MSWAP & Bond Market Association Municipal Swap/State \& Local Government 20-yr GO bond spread & 10 \\
NACMM & NACM Survey of Credit Managers: Credit Manager's Index & 10 \\
NCL & Commercial Bank Noncurrent/Total Loans & 10 \\
NFC & Nonfinancial business debt outstanding/GDP & 12 \\
OEQ & S\&P 500, S\&P 500 mini, NASDAQ 100, NASDAQ mini Open Interest & 12 \\
OINT & 3-mo. Eurodollar, 10-yr/3-mo. swap, 2-yr and 10-yr Treasury Open Interest & 12 \\
PENS & Total Assets of Pension Funds/GDP & 12 \\
RATELIQ & CME Eurodollar/CBOT T-Note Futures Market Depth & 12 \\
REIT & Total REIT Assets/GDP & 12 \\
REPO & Fed Funds/Overnight Treasury Repo rate spread & 11 \\
REPOGCF & GCF Treasury/MBS Repo rate spread & 11 \\
REPOGR & Repo Market Volume (Repurchases+Reverse Repurchases of primary dealers) & 11 \\
RRE & FRB Senior Loan Officer Survey: Tightening Standards on RRE Loans & 10 \\
RTA & Commercial Bank Real Estate Loans/Total Assets & 12 \\
RTERM & 3-mo./1-wk Treasury Repo spread & 11 \\
SBD & Total Assets of Broker-dealers/GDP & 12 \\
SMALL & NFIB Survey: Credit Harder to Get & 10 \\
SPCILARGE & FRB Senior Loan Officer Survey: Increasing spreads on Large C\&I Loans & 10 \\
SPCISMALL & FRB Senior Loan Officer Survey: Increasing spreads on Small C\&I Loans & 10 \\
SPOVX & CBOE Crude Oil Volatility Index, OVX & 10 \\
SPR210 & 10-yr/2-yr Treasury yield spread & 11 \\
SPR23M & 2-yr/3-mo. Treasury yield spread & 11 \\
STA & Commercial Bank Securities in Bank Credit/Total Assets & 12 \\
STKGR & New US Corporate Equity Issuance (Relative to 12-mo. MA) & 12 \\
STLOC & Federal, state, and local debt outstanding/GDP & 12 \\
SWAP10 & 10-yr Interest Rate Swap/Treasury yield spread & 11 \\
SWAP2 & 2-yr Interest Rate Swap/Treasury yield spread & 11 \\
SWAP3M & 3-mo. Overnight Indexed Swap (OIS)/Treasury yield spread & 11 \\
TABS & Total Assets of ABS issuers/GDP & 12 \\
TED & 3-mo. LIBOR/CME Term SOFR-Treasury spread & 11 \\
TERM & 1-yr./1-mo. LIBOR/CME Term SOFR spread & 11 \\
USD & Advanced Foreign Economies Trade-weighted US Dollar Value Index & 11 \\
VIX & CBOE Market Volatility Index VIX & 11 \\
VOL1 & 1-mo. BofAML Option Volatility Estimate Index & 11 \\
VOL3 & 3-mo. BofAML Swaption Volatility Estimate Index & 11 \\
W500 & Wilshire 5000 Stock Price Index & 12 \\
\end{longtable}
}

\end{document}


\title{Online Appendix for \emph{Machine-learning Growth at Risk}\thanks{The views expressed herein are those of the authors and should not be
attributed to the IMF, its Executive Board, or its management. }\\
}
\date{May 2025}

\author{ \small{Tobias Adrian} \thanks{Monetary and Capital Markets Department, International Monetary Fund, tadrian@imf.org.} \and
        \small Hongqi Chen\thanks{College of Finance and Statistics, Hunan University, hongqichen@hnu.edu.cn.}  \and 
        \small Max-Sebastian Dovì\thanks{African Department, International Monetary Fund, mdovi@imf.org.} \and 
        \small Ji Hyung Lee\thanks{{Department of Economics, University of Illinois Urbana-Champaign, jihyung@illinois.edu.}}
        }



\maketitle

\renewcommand{\thesection}{\Alph{section}}
\numberwithin{equation}{section}
\setcounter{section}{0}

Section \ref{Section-Upside-OA} presents the results for the analysis in the main text for $\tau = 0.95$. Section \ref{Section-GDP} presents the results for the analysis in the main text replacing IP growth with GDP growth.

\section{Drivers of upside risk\label{Section-Upside-OA}}

We repeat the analyses in the main paper for the upper quantile $\tau = 0.95$. 

\begin{figure}[H]
    \centering
    \includegraphics[width=1.1\linewidth]{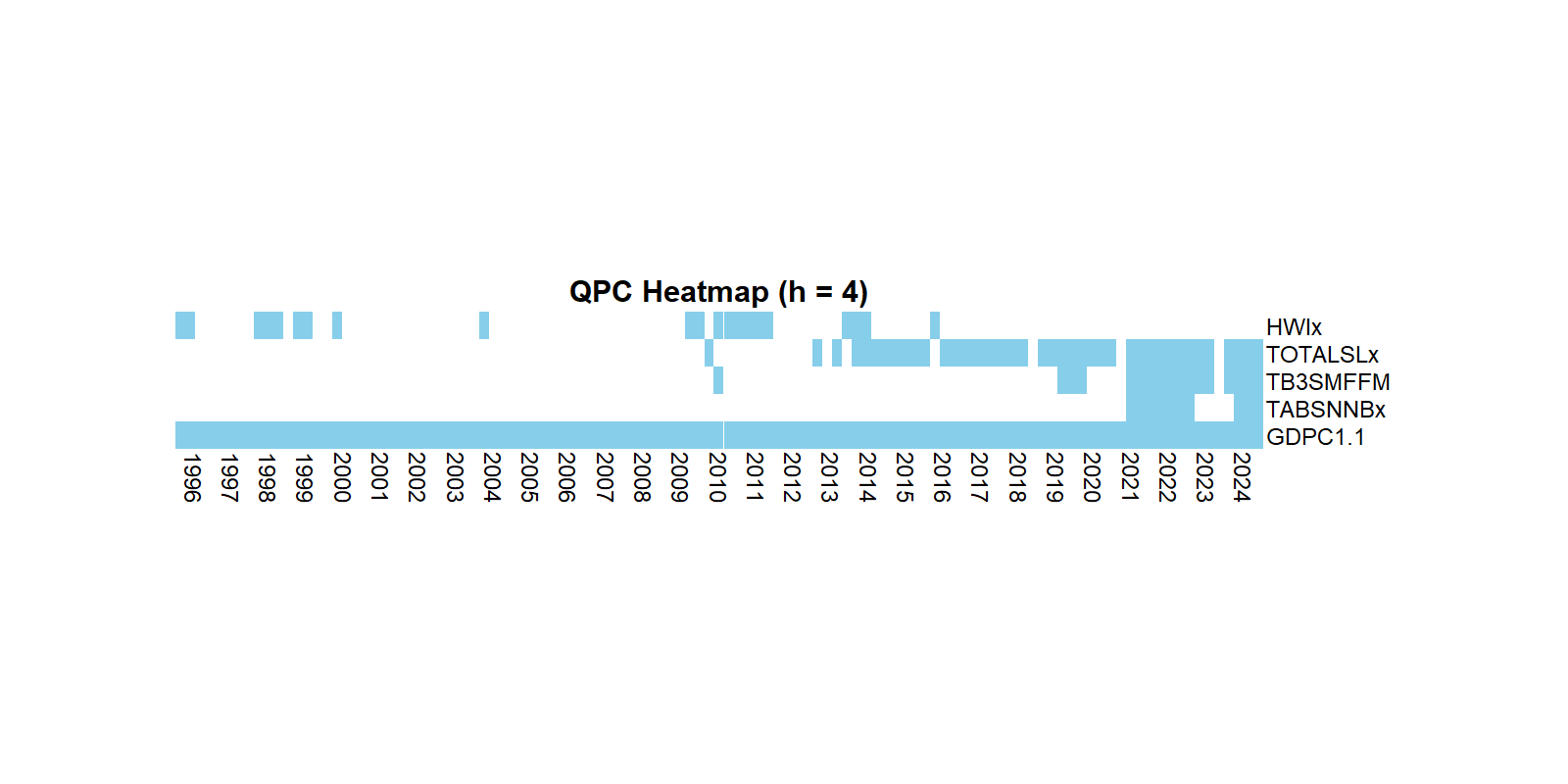}     
    \caption{Predictors selected in at least four consecutive quarters over time, $\tau = 0.95$.}
    \label{Figure-QPCS-Select-Heat-0.95}
    \footnotesize
    \begin{note*}
        \textit{Variables selected for at least 12 consecutive months. Each row represents a selected variable, and blue cells indicate periods of selection. HWIx: help-wanted index; TOTALSLx: total consumer credit outstanding, deflated by core PCE; TB3SMFFM: three-month Treasury rate minus Federal Funds Rate; TABSNNBx: real nonfinancial noncorporate business sector assets deflated by implicit price deflator; GDPC1.1: lag of GDP growth.}
    \end{note*}
\end{figure}

The heat map of systematically selected predictors for $\tau = 0.95$ is shown in Figure \ref{Figure-QPCS-Select-Heat-0.95}. We find that the main drivers of upside risk (after controlling for the lag of industrial production) to GDP can be categorised into the same four groups as for downside risk: capacity utilisation (CUMFNS, BUSINVx), labour-market indicators (PAYEMS, SRVPRD, USFIRE, CES0000000, AWHMAN), housing-related indicators (PERMIT, PERMITNE, PERMITMW, PERMITS, PERMITW), and financial indicators (TB3SMFFM, TB6SMFFM, T1YFFM, T5YFFM, T10YFFM, AAAFFM, BAAFFM). However, while the broad categories are the same, the predictors chosen are different. For instance, whereas spreads at very short-term maturities mattered for downside risk, spreads at longer maturities matter for upside risk. Moreover, indicators of financial market volatility, such as the VIX, matter for downside but not upside risk.

\begin{figure}[H]
    \centering
    \includegraphics[width=1\linewidth]{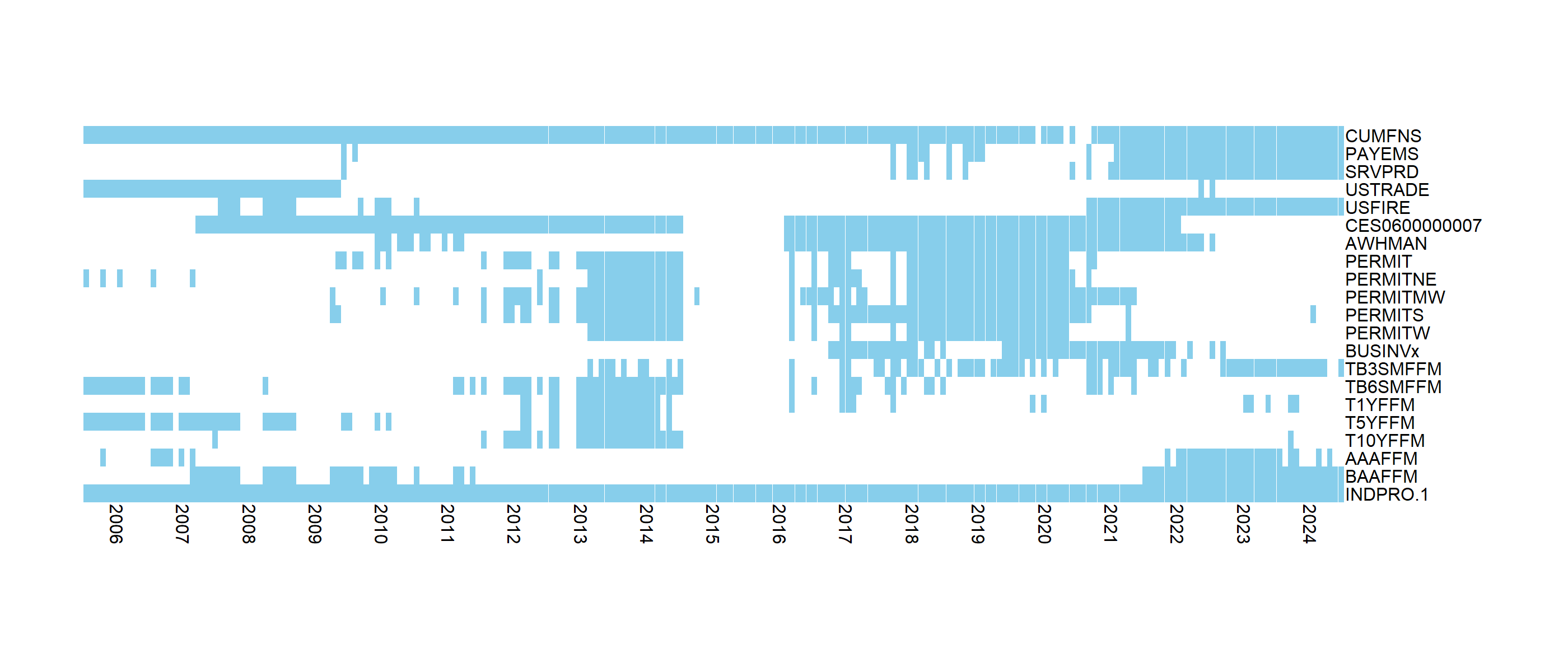}
    \caption{Frequency of variable selection for $\tau = 0.95$ }
    \label{Figure-QPCS-Select-Heat-0.95-OA}
    \footnotesize
    \begin{note*}
    \textit{Variables selected for at least 12 consecutive months. Each row represents a selected variable, and blue cells indicate periods of selection. CUMFNS: capacity utilisation; PAYEMS: total nonfarm employees; SRVPRD: service-producing employees; USTRADE: retail-trade employees; USFIRE: finance employees; CES0600000007: average weekly hours (goods producing); AWHMAN: average weekly hours (manufacturing); PERMIT: new private housing permits; PERMITNE: new private housing permits (northeast); PERMITMW: new private housing permits (midwest); PERMITS: new private housing permits (south); PERMIT: new private housing permits (west); BUSINVx: total business inventories; TB3SMFFM: three-month Treasury rate minus Federal Funds rate; TB6SMFFM: six-month Treasury rate minus Federal Funds rate; T1YFFM: one-year Treasury rate minus Federal Funds rate;  T5YFFM: one-year Treasury rate minus Federal Funds rate; T10YFFM: ten-year Treasury rate minus Federal Funds rate; AAAFFM: Moody's Aaa corporate bond rate minus Federal Funds rate; BAAFFM: Moody's Baa corporate bond rate minus Federal Funds rate; INDPRO.1: lag of industrial production..}

    \end{note*}

\end{figure}

\begin{figure}[H]
    \centering
    \begin{tabular}{cc}
        \subfloat[CUMFNS]{
            \includegraphics[width=0.45\linewidth]{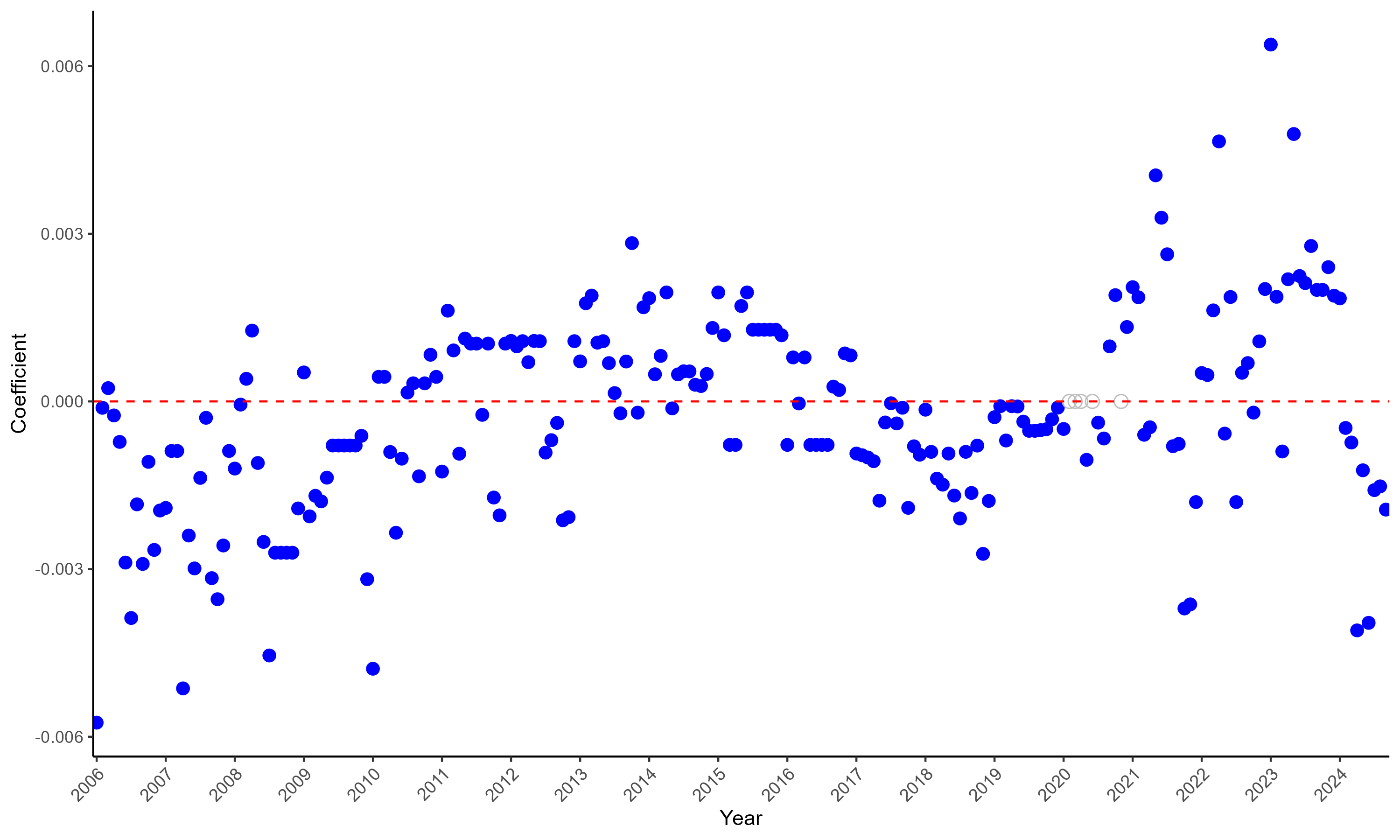}
        } &
        \subfloat[PAYEMS]{
            \includegraphics[width=0.45\linewidth]{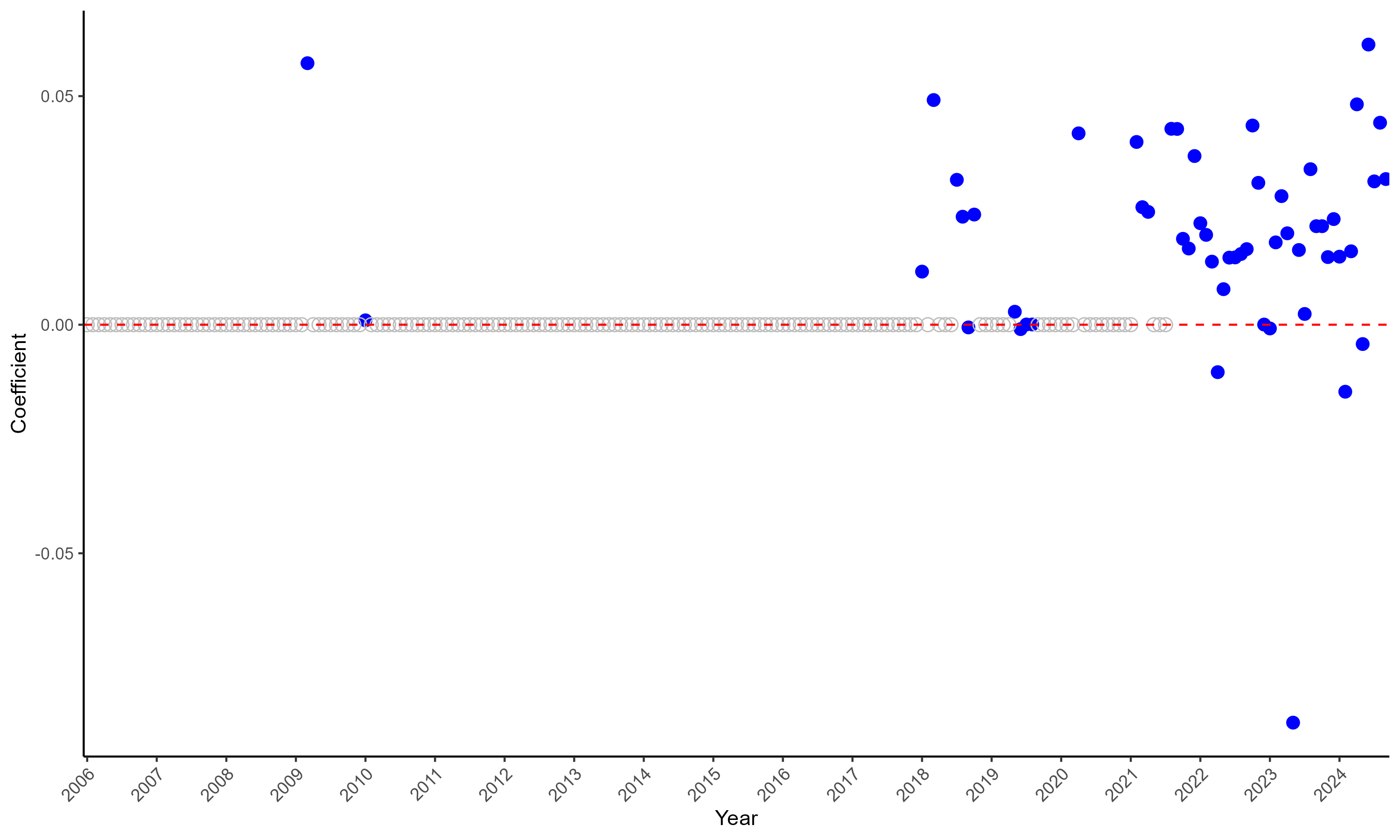}
        } \\
        \subfloat[AWHMAN]{
            \includegraphics[width=0.45\linewidth]{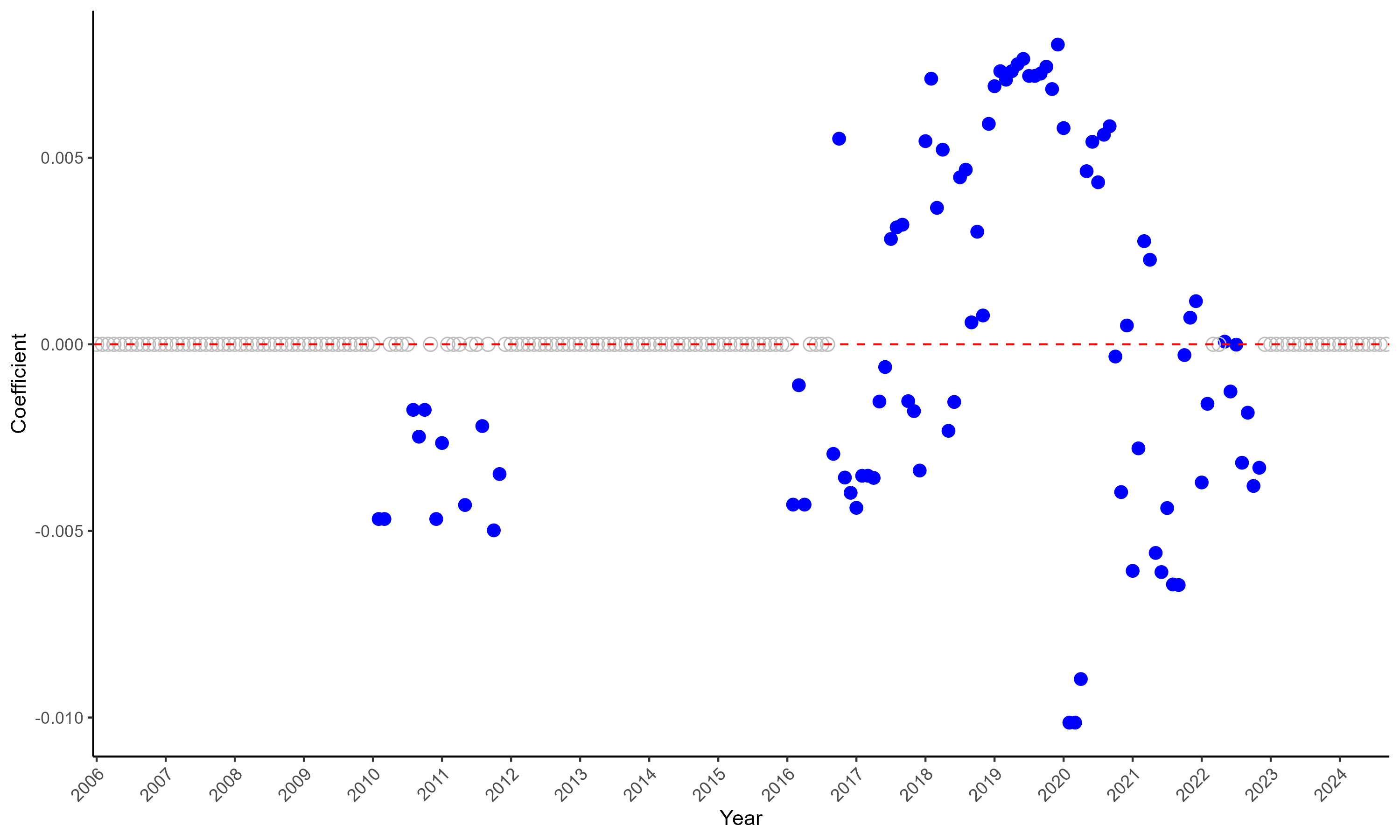}
        } &
        \subfloat[PERMIT]{
            \includegraphics[width=0.45\linewidth]{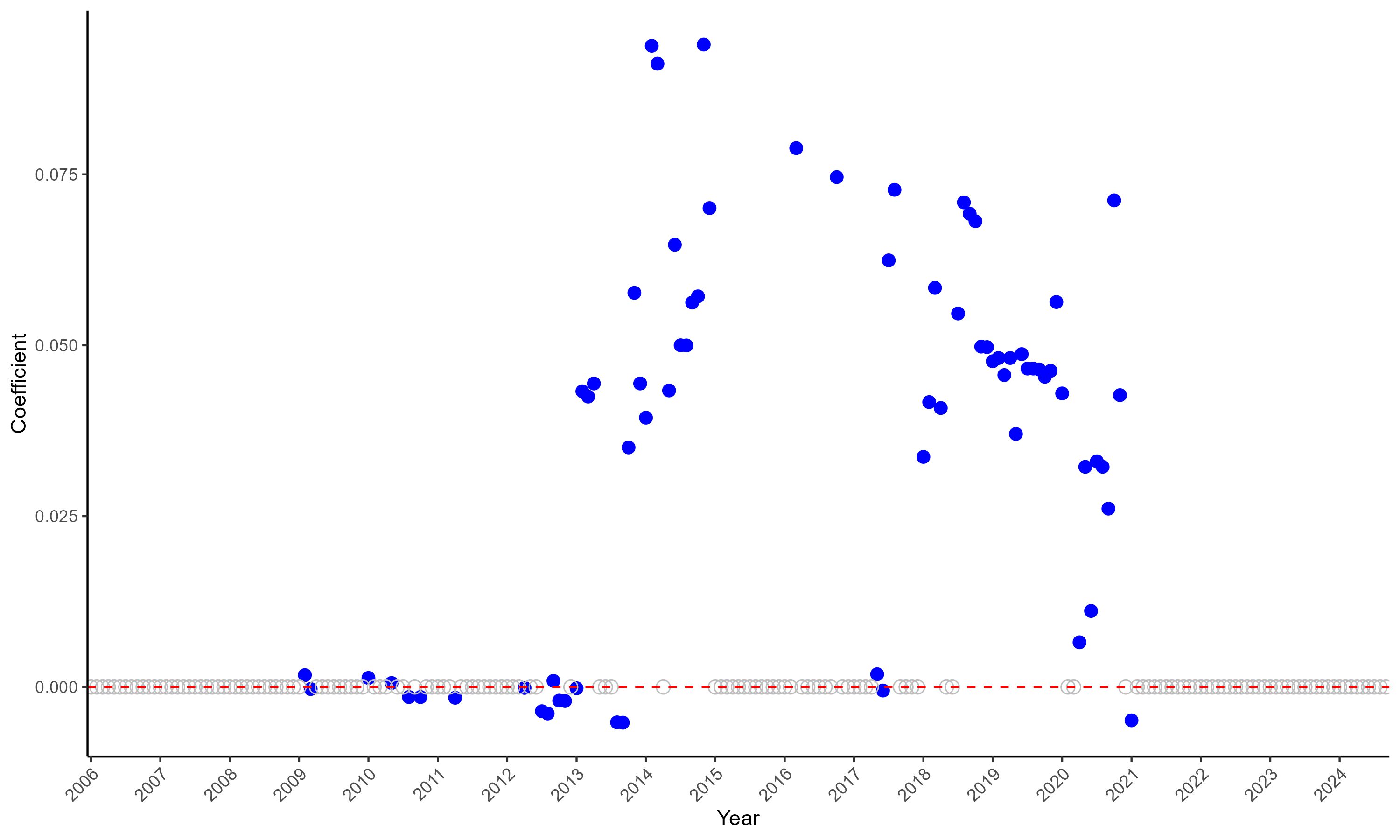}
        } \\
        \subfloat[T1YFFM]{
            \includegraphics[width=0.45\linewidth]{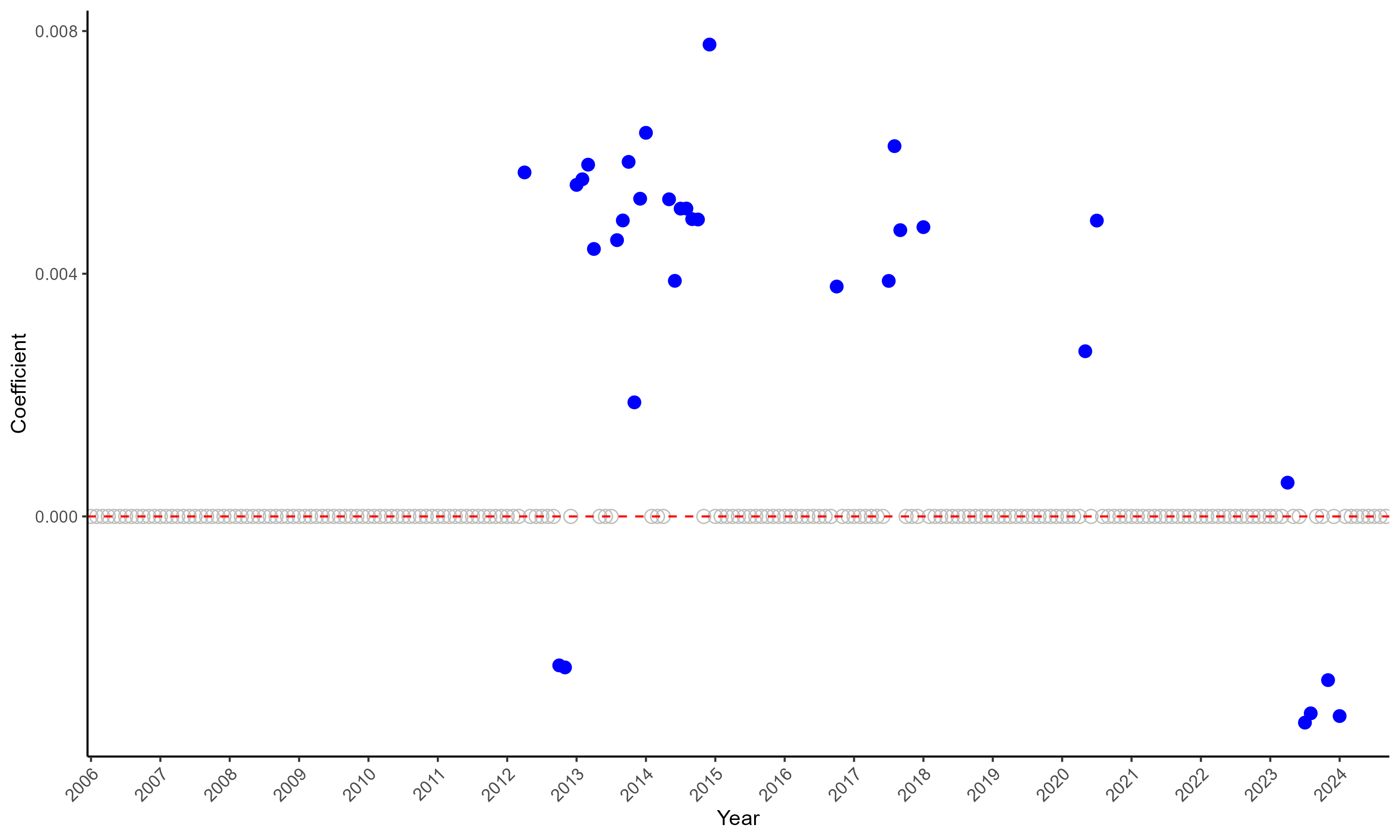}
        } & 
        \subfloat[BAAFFM]{
            \includegraphics[width=0.45\linewidth]{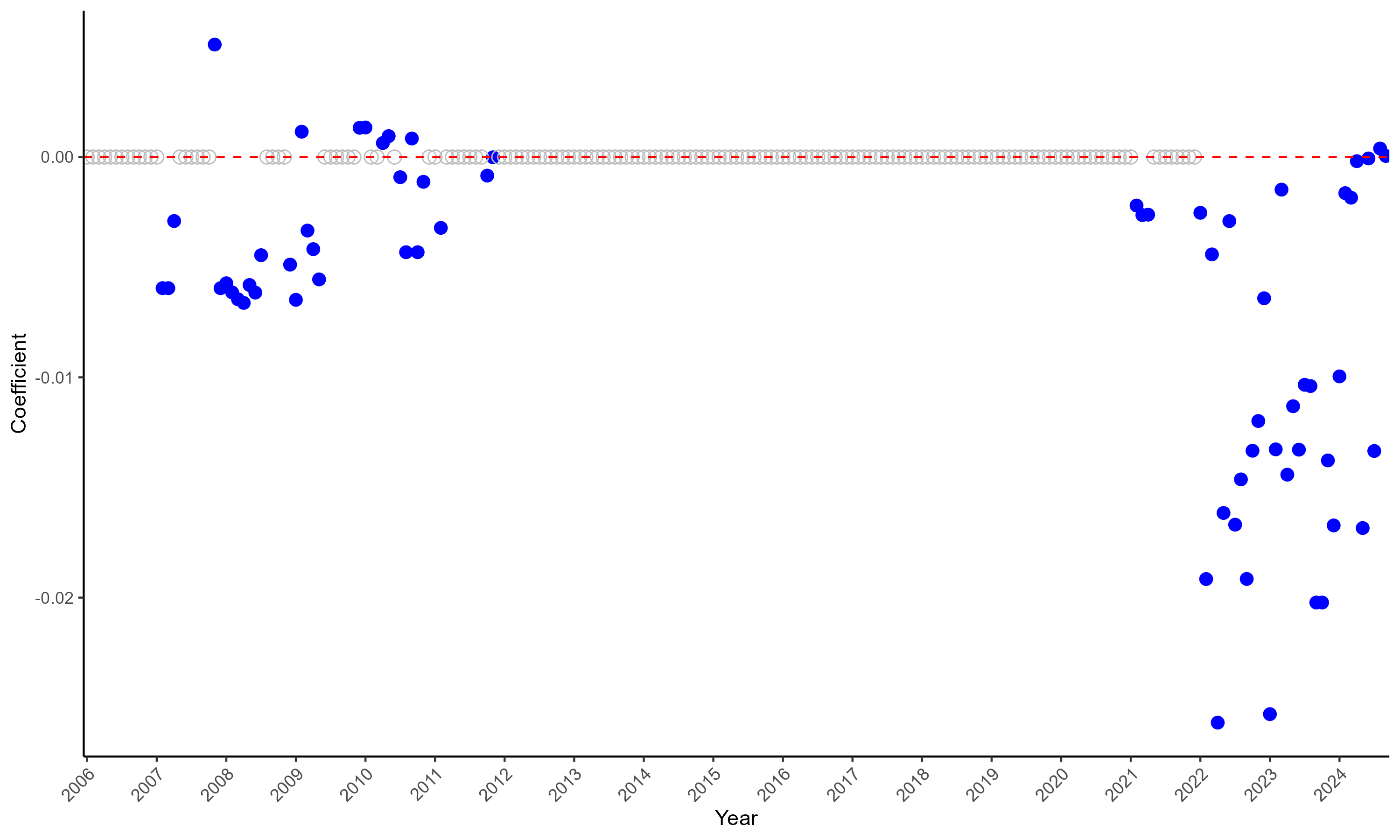}
        }\\
    \end{tabular}
    \caption{Coefficients of selected variables for $\tau = 0.95$ over time.}
    \label{Figure-QPCS-Coeffs-0.95}
    \footnotesize
    \begin{note*}
        \textit{Solid blue dots indicate periods of selection and show the corresponding value of the estimate. Hollow dots represent periods in which the variable was not selected. See Figure \ref{Figure-QPCS-Select-Heat-0.95-OA} for definitions of the acronyms.}
    \end{note*}
\end{figure}

Figure \ref{Figure-QPCS-Coeffs-0.95} shows coefficients over time of some of the selected variables in the groups described above: CUMFNS (capacity utilisation), PAYEMS (total nonfarm employees), AWHMAN (average weekly hours of production), PERMIT (new private housing permits), BAAFFM (spread of Baa corporate bonds rate to the Federal Funds rate), and T1YFFM (spread of one-year treasury yield to the Federal Funds rate). 

The first panel shows that the sign of the effect of capacity utilisation on `growth tailwinds' varies substantially over time, making it difficult to attach economic interpretation to this coefficient. By contrast, the second panel shows that increases in hours worked positively affect upside risk to growth, especially in the period 2022-2024, likely reflecting the robust US labour market coming out of the Covid-19 pandemic. The third panel shows that increases in average hours worked have had an ambiguous effect on uspide risk to growth over time. The fourth panel shows that increases in housing permits--possibly reflecting an optimistic economic outlook--between the end of the GFC and the start of the Covid-19 pandemic have improved upside growth risk. The fifth panel shows that increases in the corporate bond spreads decrease upside growth risk. The sixth panel shows that increases in the spread of one-year Treasury-bill rates to the Federal Funds rate--which can be interpreted as yield curve steepenings--improve upside growth risk. 

We again note that the results in Figure \ref{Figure-QPCS-Coeffs-0.95} confirm commonly held heuristics to assess uspide growth. For instance, a strong labour market, an optimistic outlook on the housing market, as well as yield-curve steepenings are predictive of upside growth risk. We also note that similarly to the case of downside growth risk, yield curve steepenings increase upside risk, while non-governmental spreads decrease it. This suggests that both factors affect the entire predictive distribution of growth. 

\section{Application to GDP Growth \label{Section-GDP}}

We apply the same methodology in the main text to forecast quantiles of GDP growth using quarterly data taken from Fred-QD. The results confirm the qualitative findings reported in the main text: financial predictors are systematically selected for forecasting both downside and upside risk, and that the sources of risk vary over time.

\subsection{Drivers of downside risk}

 To avoid muddying our discussion by predictors that are only sporadically selected, we focus we focus on those predictors that are selected in at least four consecutive quarters. We find that predictors related to the housing market (PRFIx) and industrial production (IPNCONGD) are consistently selected in roughly the run-up to the GFC. Predictors related to the labour market (USPBS and CLAIMSx) are more consistently selected thereafter, in particularly in the wake of and following the Covid-19 pandemic. Financial predictors (TB6M3Mx, M2REAL, AAAFFM, and TLBSNNCBx) are selected throughout the entire period we consider, although somewhat more sporadically prior to the GFC.

The decreased role of residential investment in predicting downside risk post-GFC may be indicative of both the importance of the housing market in the lead-up to the GFC and the increased regulation of financial products associated with it following the GFC. The decreased role of the industrial production of durable consumer goods (IPNCONGD) is consistent with structural changes in the economy around the mid-2000s that saw the rise of the digital economy. The increased importance of financial conditions during and after the GFC is consistent with the view that there were substantial risks emanating from the financial sector, and that monetary policy played an important role in avoiding the most damaging materialisation of them. In particular, the increased importance of corporates' leverage (TLBSNNCBBDIx) in explaining downside risk to GDP is consistent with the view that the GFC was in part due to excessive buildup of leverage (see, e.g., \citet{Geanakoplos}).  The importance of labour-market predictors during the Covid-19 pandemic is consistent with the view that most of the downside risks to GDP growth emanated from large layoffs, as well as the disruptions to production caused by the inability to meet due to lockdowns.

\begin{figure}[H]
    \centering
    \includegraphics[width=1.1\linewidth]{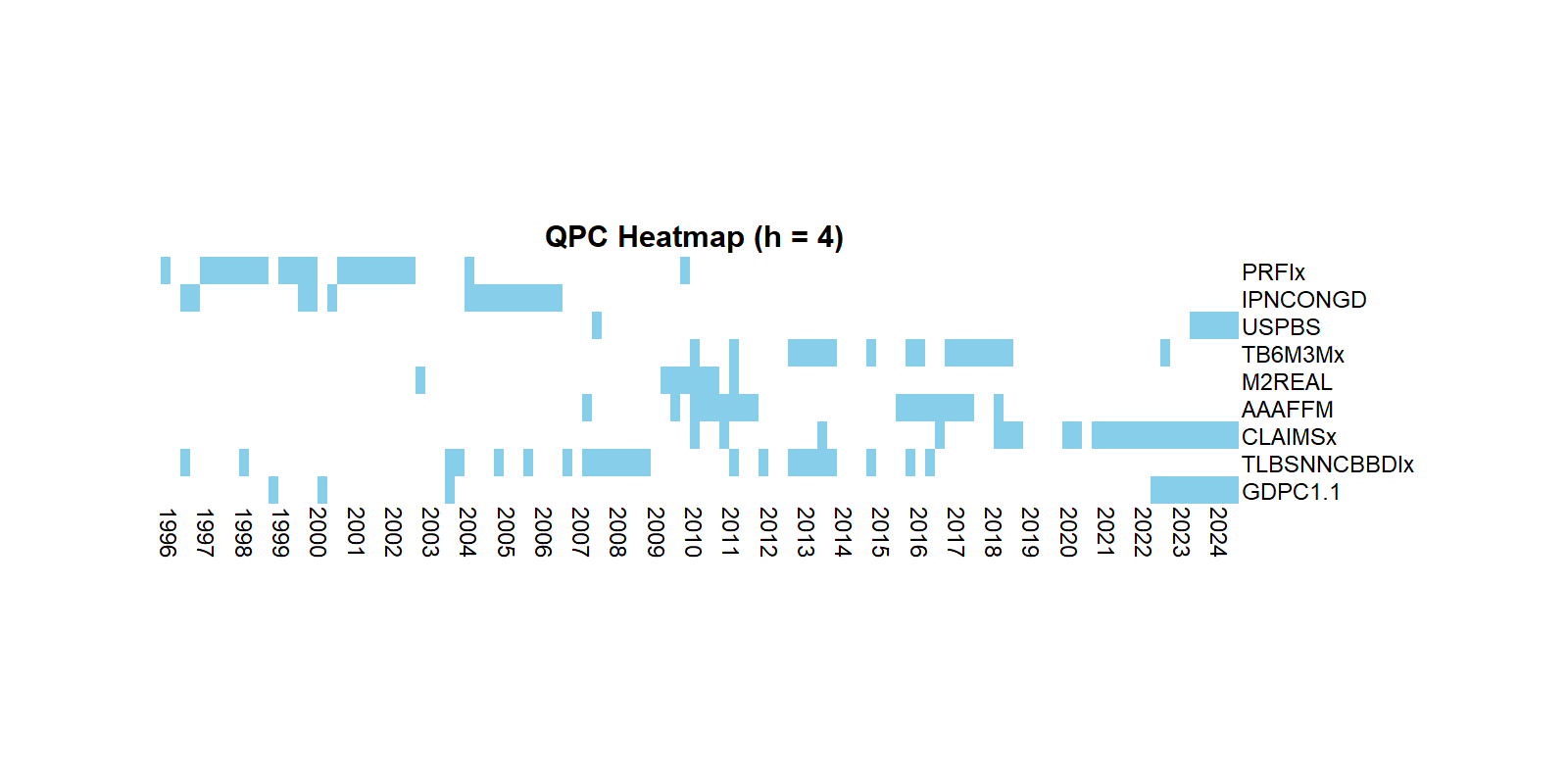}     
    \caption{Predictors selected in at least four consecutive quarters over time, $\tau = 0.05$.}
    \label{Figure-QPCS-Select-Heat-0.05}
    \footnotesize
    \begin{note*}
        \textit{Variables selected for at least 4 consecutive quarters. Each row represents a selected variable, and blue cells indicate periods of selection. PRFIx: real private fixed investment; IPNCONGD: industrial production nondurable consumer goods; USPBS: professional and business services employees; TB6M3Mx: six-month Treasury rate minus three-month Treasury rate; M2REAL: real money stock; AAAFFM: Moody's Aaa corporate bond rate minus Federal Funds rate; CLAIMSx: initial claims; TLBSNNCBBDIx: nonfinancial corporate business sector liabilities to disposable business income; GDPC1.1: lag of GDP growth.}
    \end{note*}
\end{figure}

\begin{figure}[H]
    \centering
    \begin{tabular}{cc}
        \subfloat[CLAIMSx]{
            \includegraphics[width=0.45\linewidth]{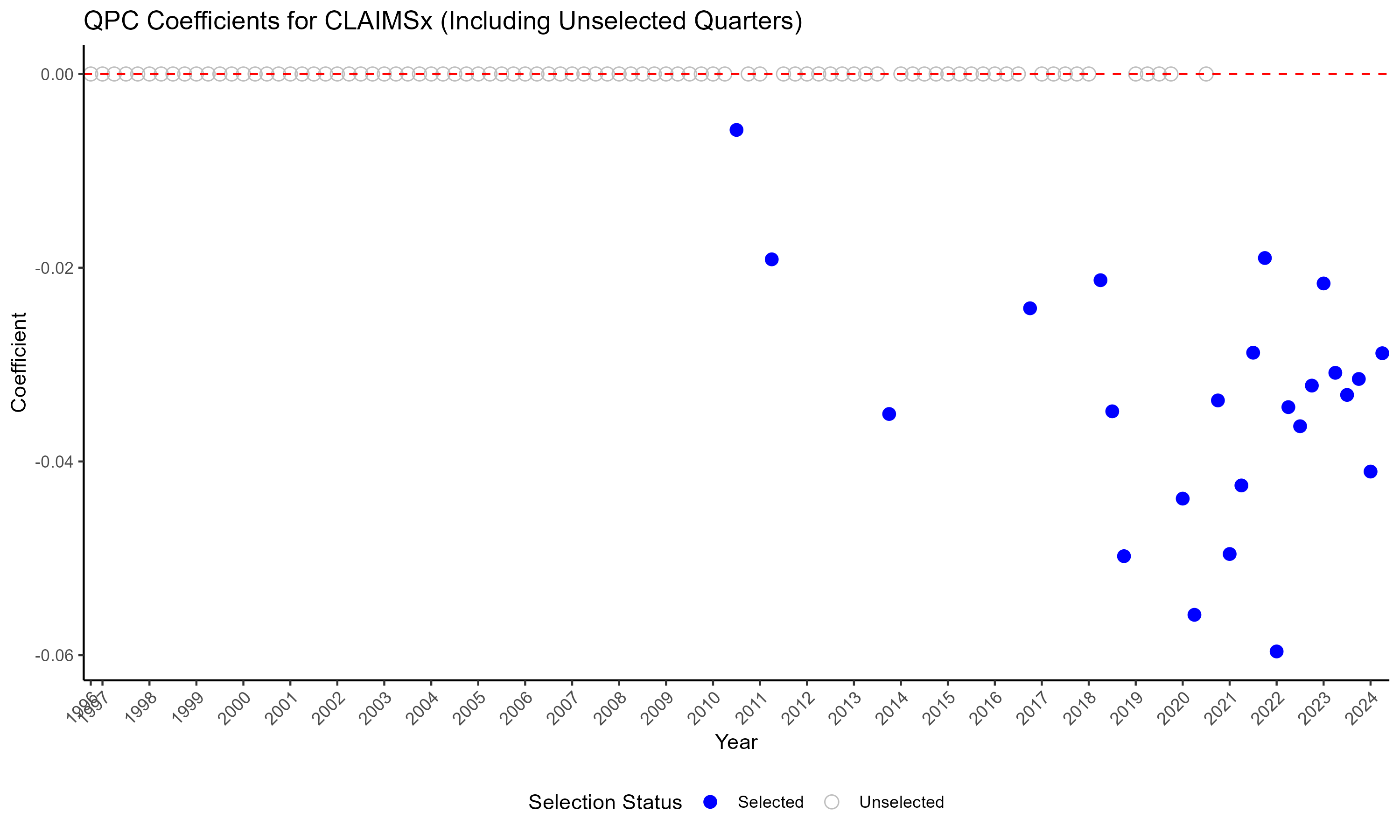}
        } &
        \subfloat[TLBSNNCBBDIx]{
            \includegraphics[width=0.45\linewidth]{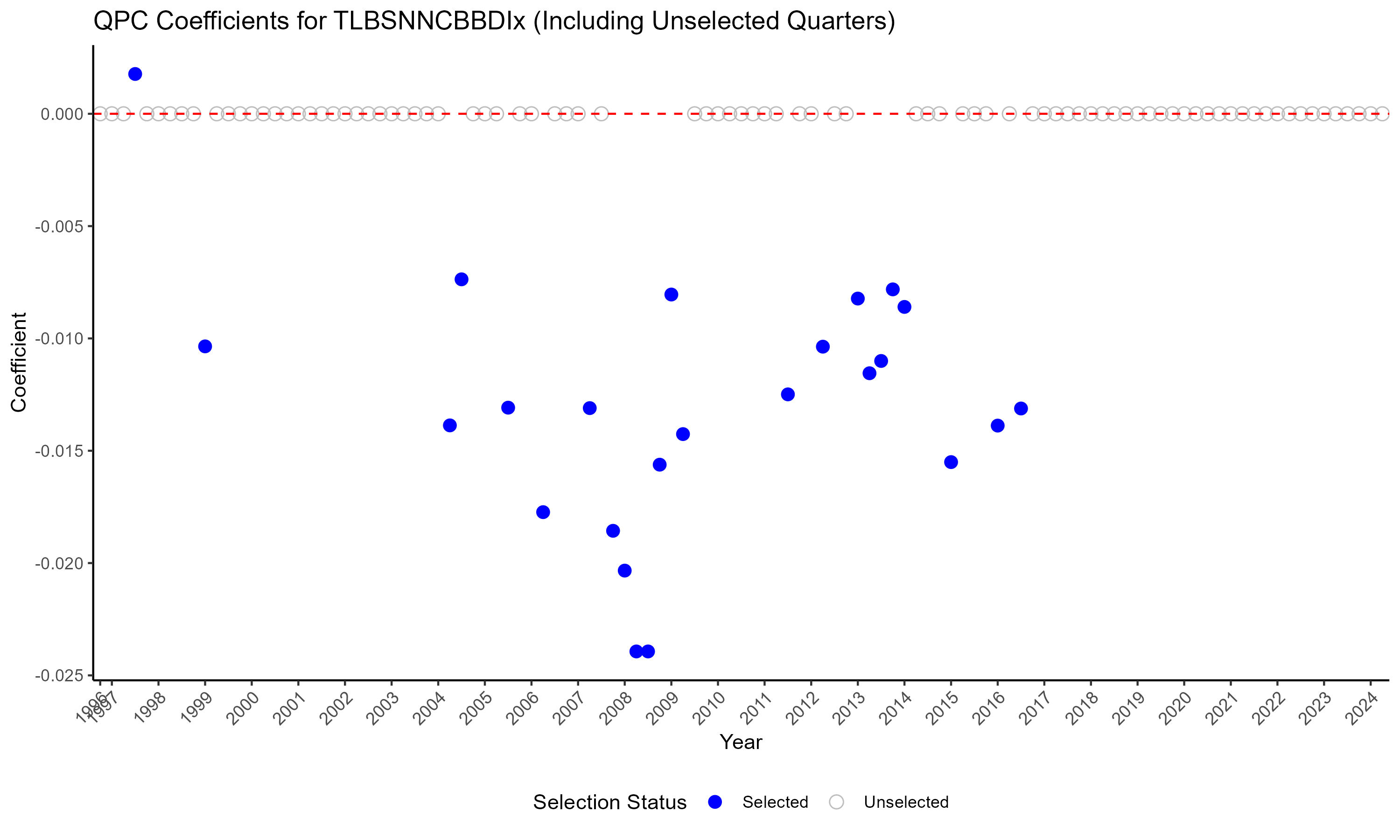}
        } \\
        \subfloat[PRFIx]{
            \includegraphics[width=0.45\linewidth]{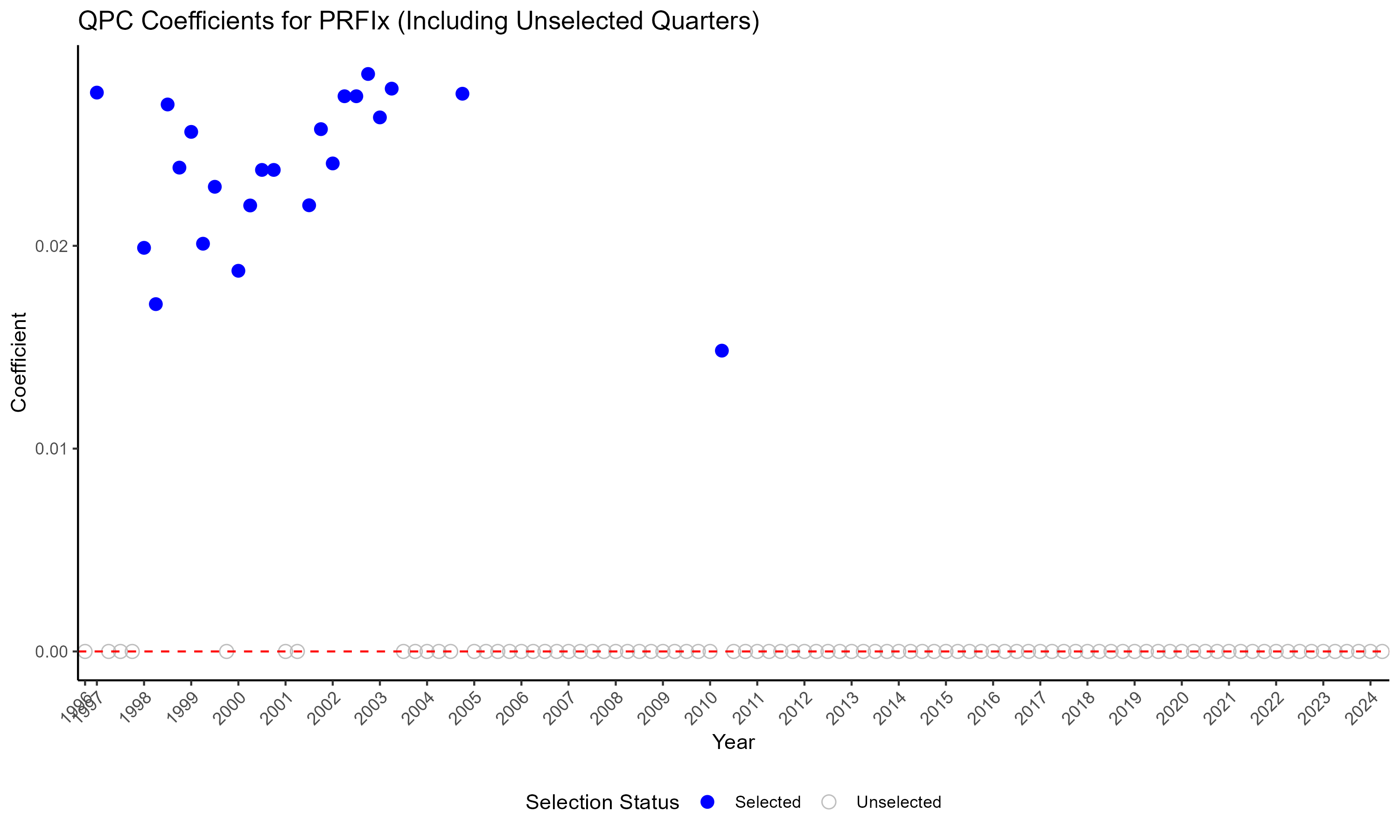}
        } &
        \subfloat[AAAFFM]{
            \includegraphics[width=0.45\linewidth]{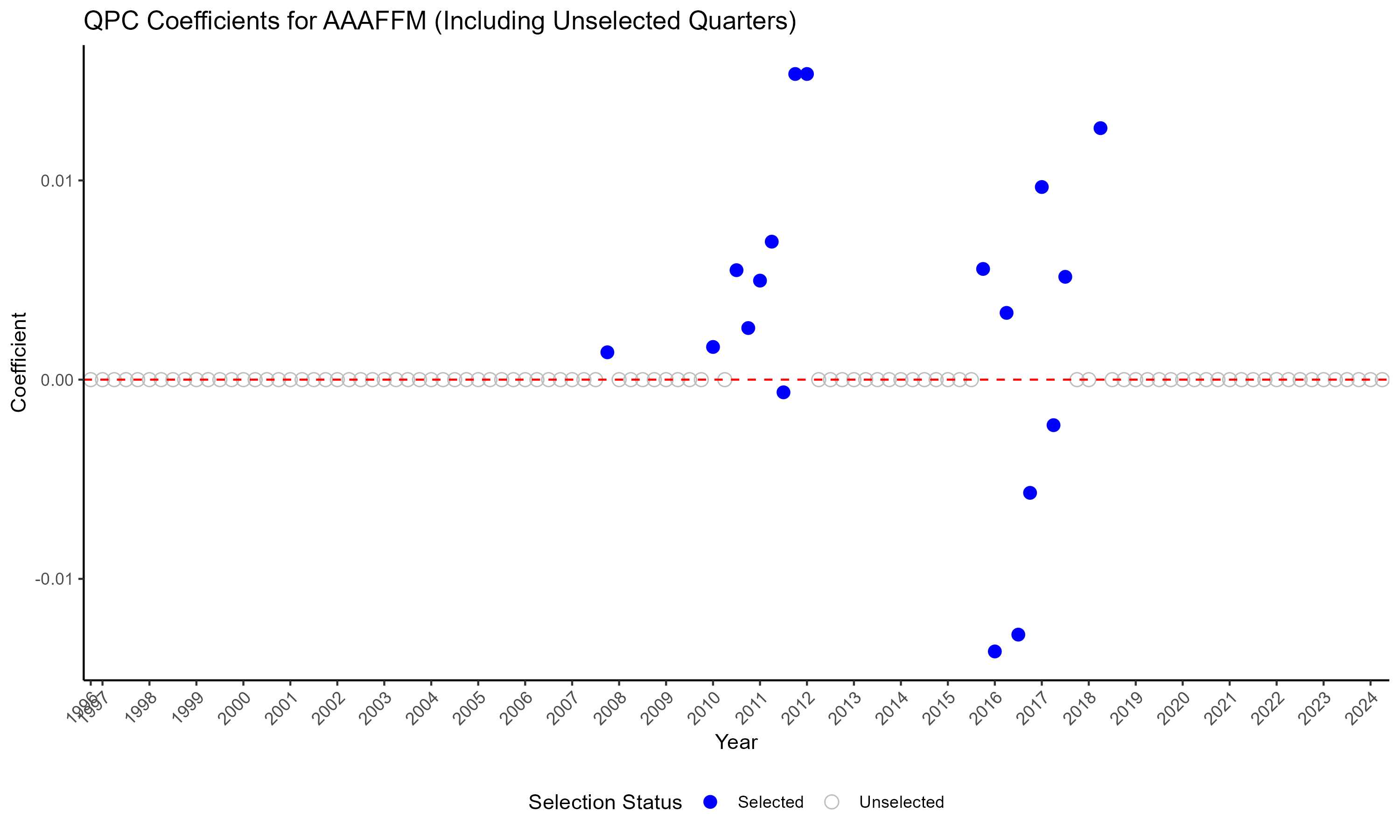}
        } \\
        \multicolumn{2}{c}{\subfloat[TB6M3Mx]{
            \includegraphics[width=0.45\linewidth]{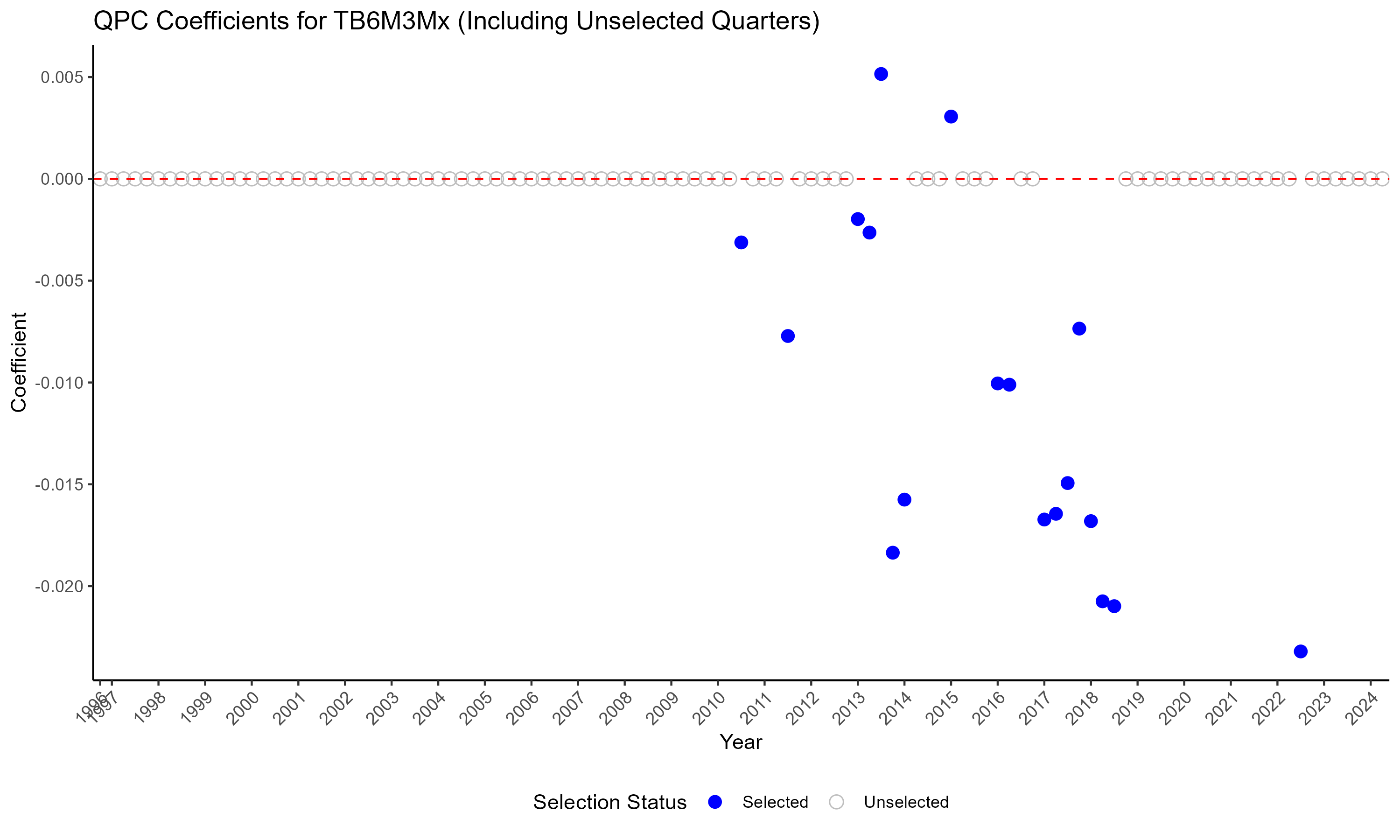}
        }
        }\\
    \end{tabular}
    \caption{Coefficients of selected variables for $\tau = 0.05$ over time.}
    \label{Figure-QPCS-Coeff-Top-5-0.05}
    \footnotesize
    \begin{note*}
        \textit{Solid blue dots indicate periods of selection and show the corresponding value of the estimate. Hollow dots represent periods in which the variable was not selected. See Figure \ref{Figure-QPCS-Select-Heat-0.05} for definitions of the acronyms.}
    \end{note*}
\end{figure}

Figure \ref{Figure-QPCS-Coeff-Top-5-0.05} shows the coefficients of the five most frequently selected predictors for $\tau = 0.05$ over time. We find that the marginal effect of higher initial unemployment claims (CLAIMs) has a negative effect on downside risk to GDP growth, consistent with economic intuition. The marginal effect of increased leverage (TLBSNNCBBDIx) also negatively impacts downside risk to GDP growth, with the effect being particularly pronounced during the Global Financial Crisis (GFC). This aligns with the view that the GFC was partly driven by an excessive buildup of leverage. The marginal effect of increased real private residential investment improves downside risk to GDP growth, although as noted above we find that this predictor stops being informative after the GFC, potentially reflecting increased regulation of the financial products associated with the housing market. 

\subsection{Drivers of upside risk}

We repeat the analysis in the previous section for $\tau = 0.95$.

Figure \ref{Figure-QPCS-Select-Heat-0.05} shows the systematically selected predictors for $\tau = 0.95$, defined (as above) as those predictors that are selected in at least four consecutive quarters. We find that fewer predictors are systematically selected for $\tau = 0.95$ than $\tau = 0.05$, suggesting that the sources of upside risks have changed more gradually over time. As documented above, lagged GDP is selected in each of the samples we consider. We find that that labour-market tightness (HWIx) is an important predictor of upside risk to GDP growth until 2016, while financial variables (TOTALSLx, TB3MFFM, and TABSNNBx) are important in the wake of the GFC. 

\begin{figure}[H]
    \centering
    \begin{tabular}{cc}
        \subfloat[GDPC]{
            \includegraphics[width=0.45\linewidth]{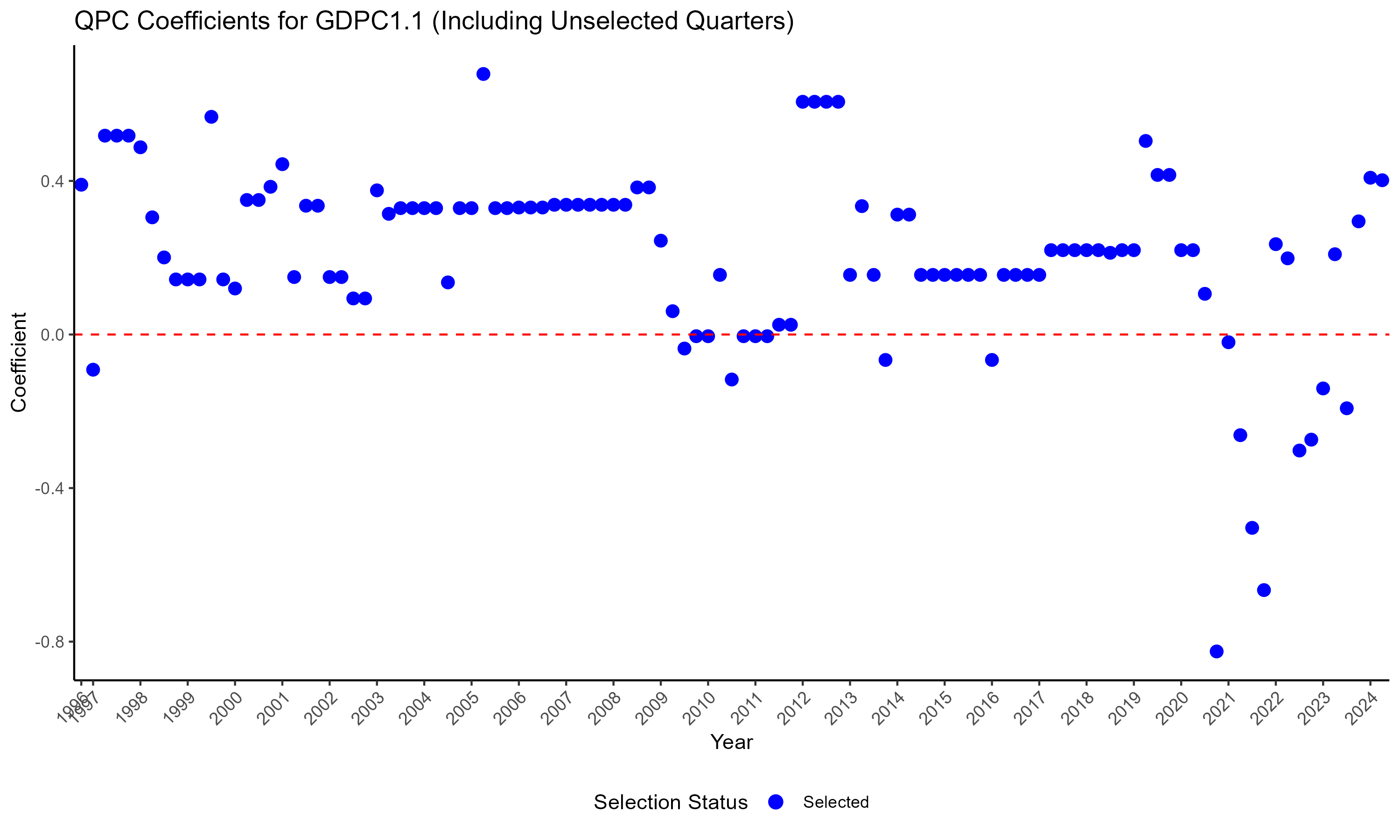}
        } &
        \subfloat[TOTALSLx]{
            \includegraphics[width=0.45\linewidth]{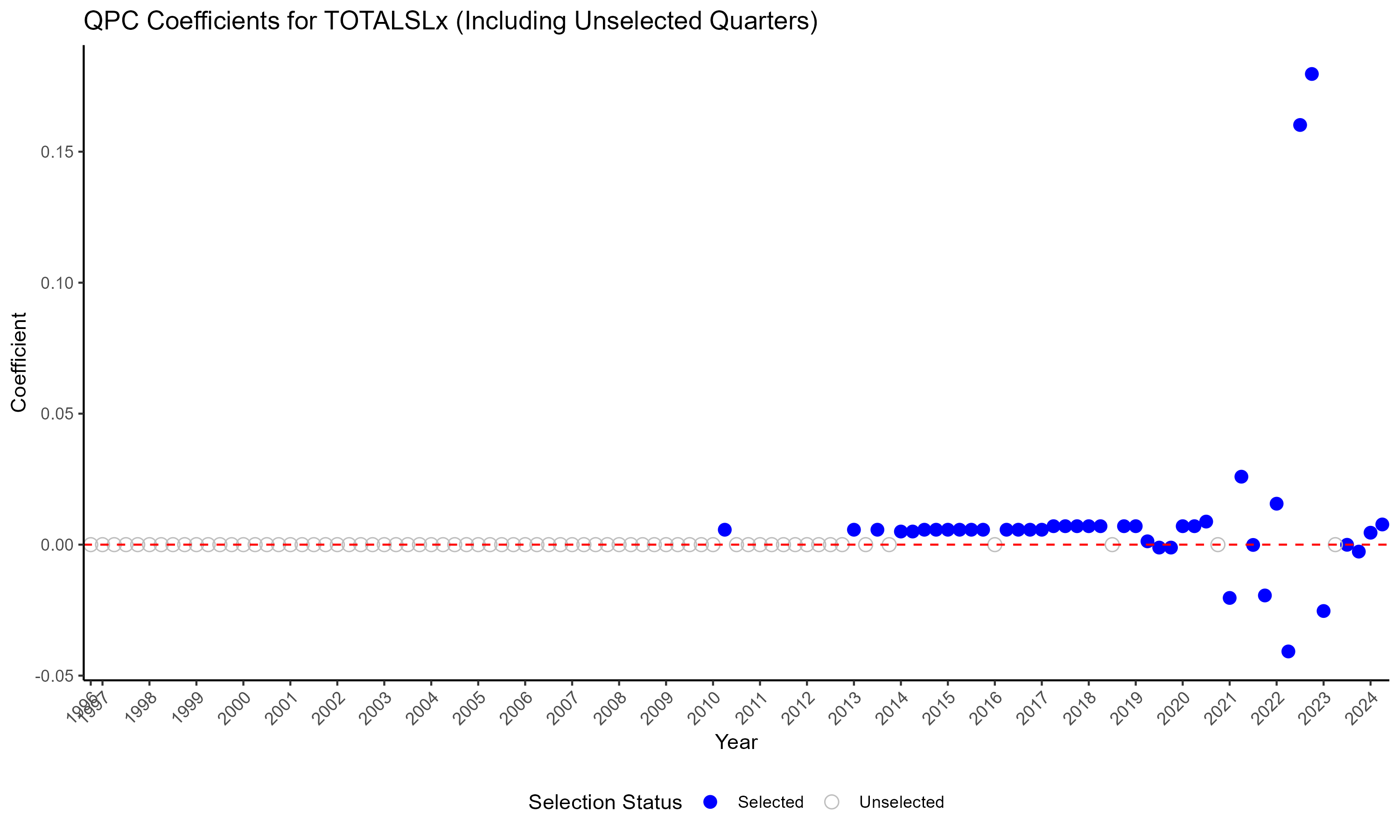}
        } \\
        \subfloat[HWIx]{
            \includegraphics[width=0.45\linewidth]{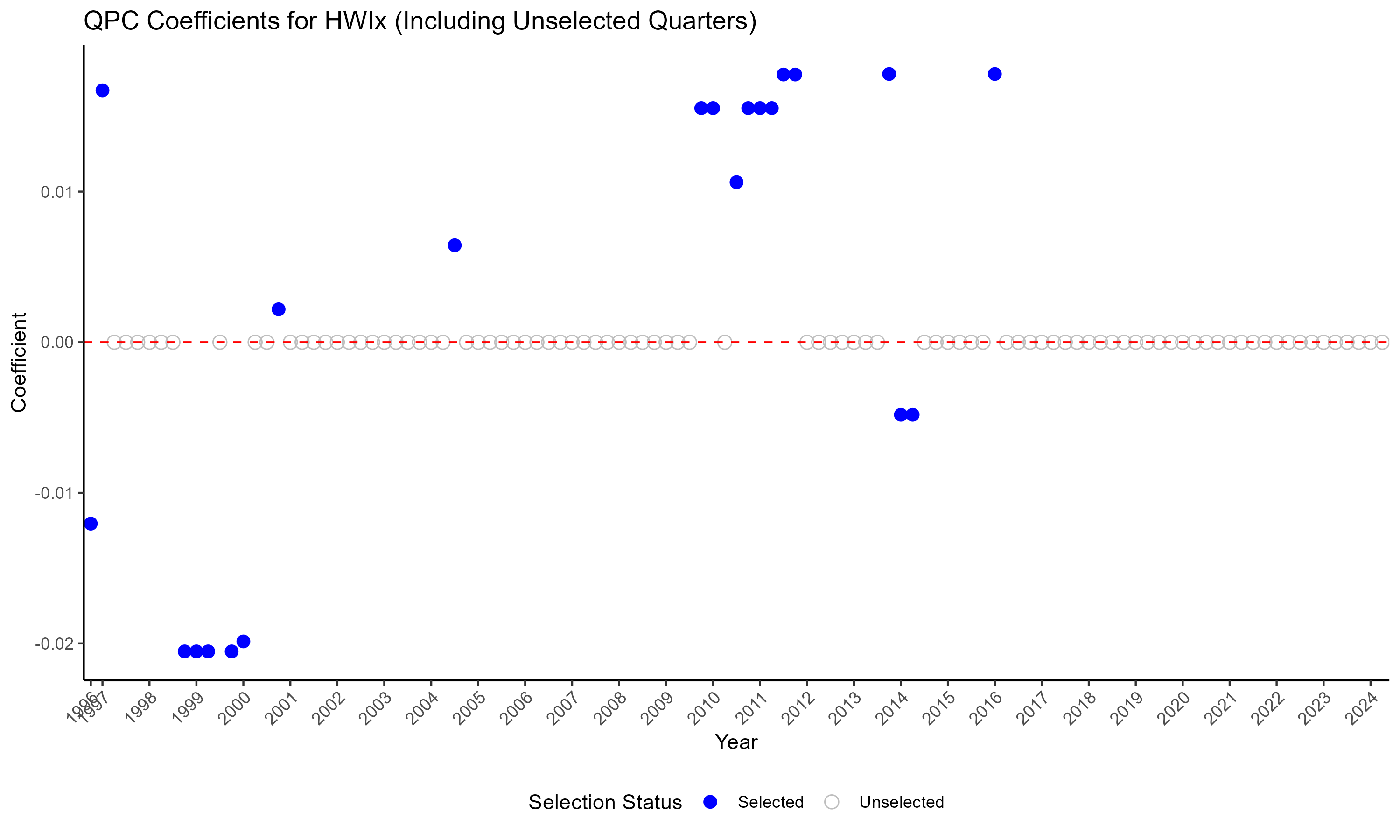}
        } &
        \subfloat[TB3SMFFM]{
            \includegraphics[width=0.45\linewidth]{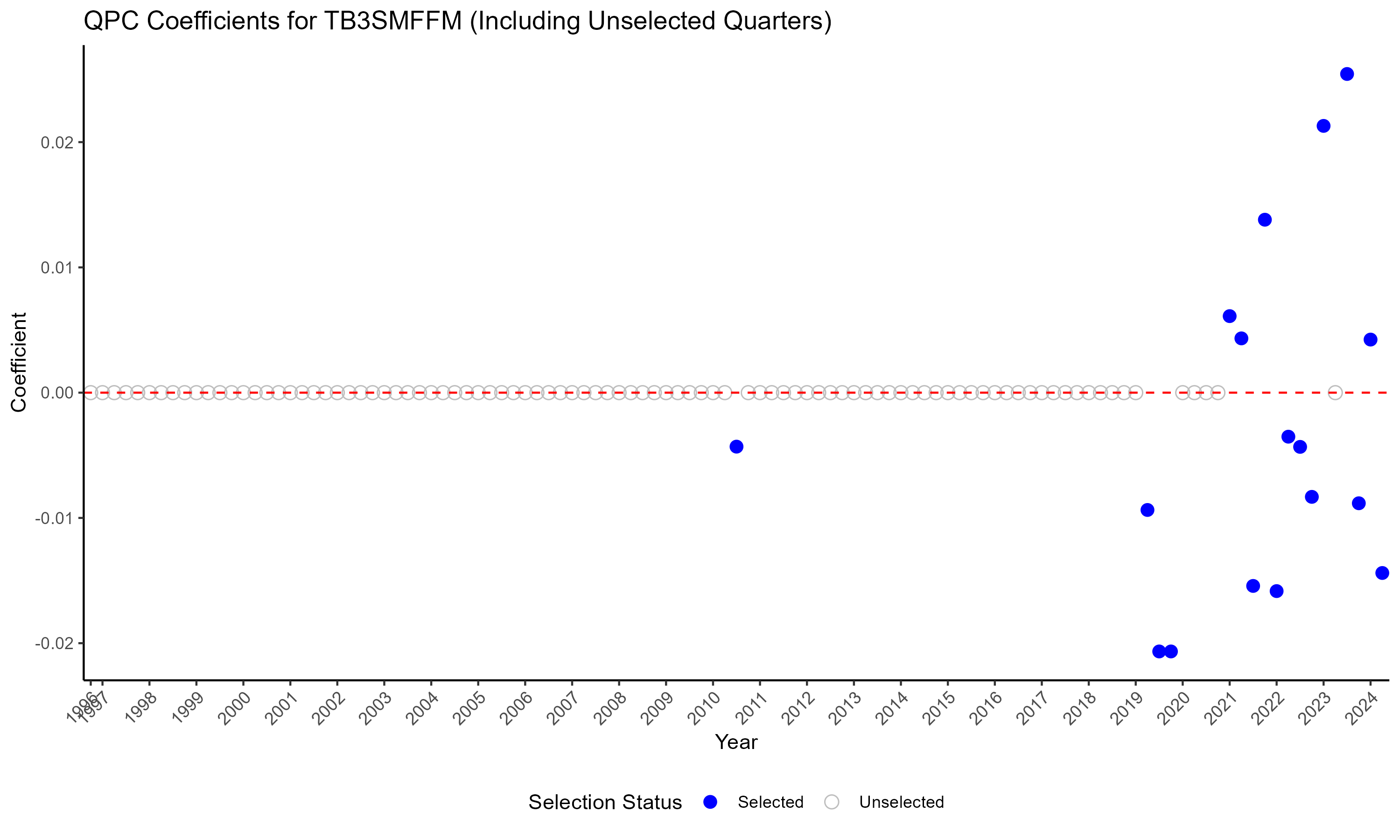}
        } \\
        \multicolumn{2}{c}{\subfloat[GFDEBTNx]{
            \includegraphics[width=0.45\linewidth]{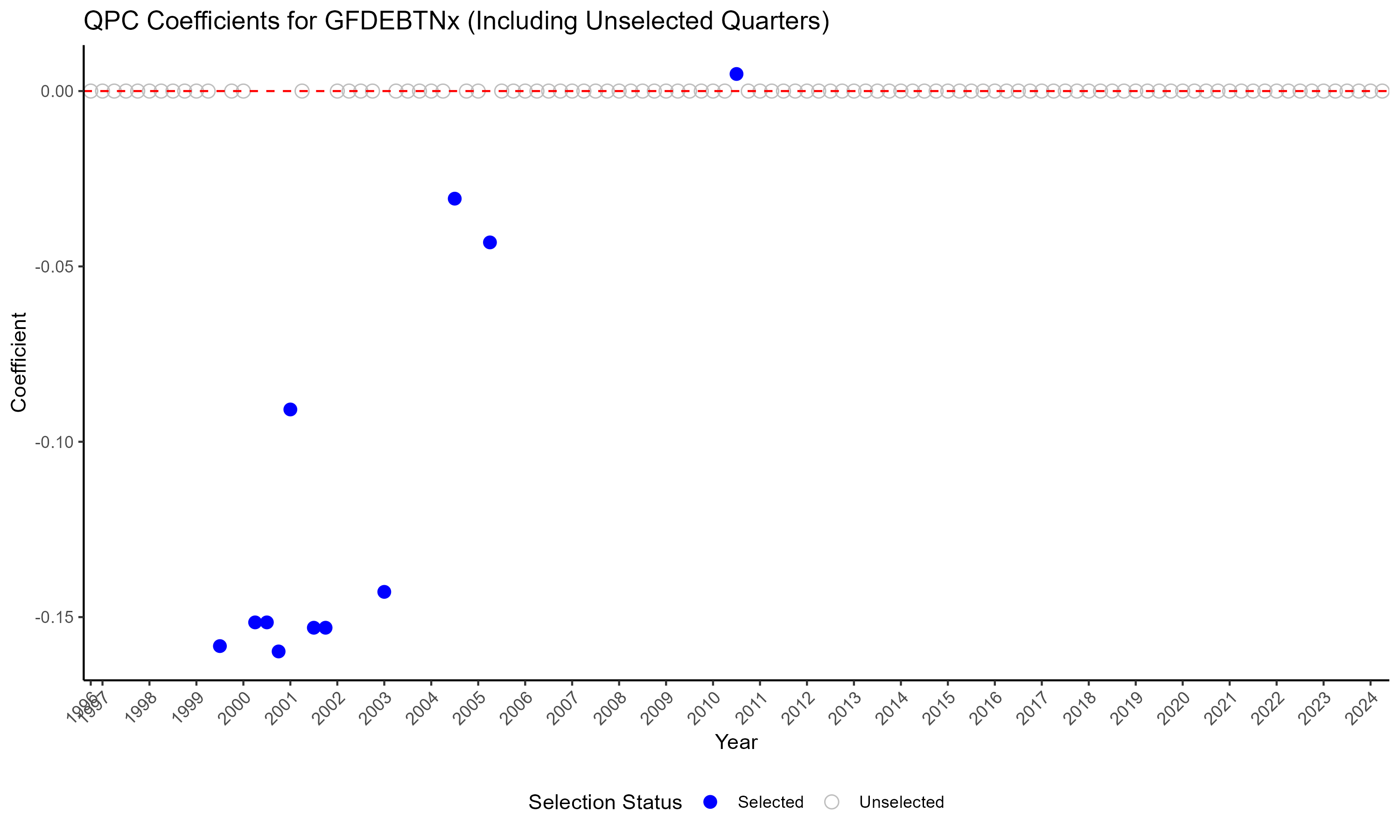}
        }
        }\\
    \end{tabular}
    \caption{Coefficients of selected variables for $\tau = 0.95$ over time.}
    \label{Figure-QPCS-Coeff-Top-5-0.95}
        \footnotesize
    \begin{note*}
        \textit{Solid blue dots indicate periods of selection and show the corresponding value of the estimate. Hollow dots represent periods in which the variable was not selected. See Figure \ref{Figure-QPCS-Select-Heat-0.95} for definitions of the acronyms.}
    \end{note*}
\end{figure}

As for the case $\tau = 0.05$, we also consider how the coefficients of the main predictors have changed over time. Figure \ref{Figure-QPCS-Coeff-Top-5-0.95}). The coefficient on lagged GDP growth is positive in most periods, suggesting that strong economic growth today improves upside risks tomorrow. 


\pagebreak

\bibliographystyle{chicago}
\bibliography{AtRisk}

\pagebreak

%% file: sections/introduction_monthly.tex
\section{Introduction}

Analyses of downside risks to real economic activity originating in the financial sector have become a crucial part of assessing the growth outlook of economies and are routinely featured in policy reports \citep{international2017global, ECB, ESRB, international2024global}. Following \citet{AdrianAER}, quantile regressions (QRs) have been widely adopted to assess such ``growth vulnerability'' -- also referred to as ``Growth-at-Risk'' or GaR. QRs are well-suited to estimate the sensitivity of quantiles of future growth to macrofinancial conditions for two key reasons. First, QRs do not impose parametric assumptions on the distribution of growth. Second, the linearity of QRs offers clear economic interpretability, a feature that is often absent in models that feature time-varying volatility such as GARCH and certain machine-learning (ML) methods including model averaging, regression trees, random forests, and neural networks.

One key limitation of QRs is that they only allow for a small number of predictors, particularly in the tail quantiles. As a result, most GaR analyses feature a single aggregated financial conditions index (FCI) that is constructed from individual financial series. However, this approach suffers from a series of drawbacks. First, it may reduce forecasting accuracy, as aggregated FCIs may not aggregate financial information in the most relevant fashion to forecast quantiles of growth \citep{Sokol}.\footnote{The large literature on mean forecasting macroeconomic data has shown that relevant predictive information can be spread across several different variables \citep{KockChapter, banbura2010large, Dalibor, medeiros2021forecasting, giannone2015prior}, and recent evidence indicates that this also applies to quantile forecasting \citep{carriero2022nowcasting, pruser2024nonlinearities, clark2024investigating, clark2023tail, carriero2024specification}.} Second, relying on aggregate FCIs can reduce interpretability, thereby complicating the derivation of clear policy implications. For instance, while a tightening of the National FCI (NFCI) has been shown to increase the likelihood of lower GDP growth, it remains unclear which specific components of the NFCI drive this result at any given time. Third, the inability to control for a comprehensive set of other macroeconomic predictors raises the concern that the QR coefficients on FCIs may incorrectly attribute the effect of non-financial variables on the conditional quantiles to the FCI \citep{plagborg2020growth}. Fourth, some FCIs--including the NFCI--are estimated using smoothing techniques such as Kalman filtering, which may introduce generated regressor problems as well as look-ahead bias \citep{amburgey2023real}.

In this paper, we employ ML methods to simultaneously consider the effect of a large set of macrofinancial predictors on downside growth risks, and hence address the shortcomings of assessing GaR using aggregated FCIs. We furthermore show how these methods can be used to construct sector-specific indices, such as FCIs, that predict downside risk while adjusting for the influence of correlated variables in other sectors.

The first contribution of this paper is the application of ML methods to identify key financial predictors of growth vulnerabilities as measured by industrial production (IP) growth. Our paper is hence related to recent work by \citet{pruser2024nonlinearities} and \citet{carriero2024specification}, but our focus differs in at least three ways. First, we focus on the 0.05 quantile as this is the threshold commonly used in policy analyses, whereas \citet{carriero2024specification} center their analysis on the 0.1 quantile. Second, and relatedly, we target quantiles of IP growth at the monthly frequency rather than GDP growth at the quarterly frequency because, as our simulations show, selection-based ML methods perform poorly in tail quantiles with limited observations.\footnote{We note that the correlation between IP growth and GDP growth is 0.75 in the sample we consider, suggesting that IP growth is a good proxy for GDP growth.} Nevertheless, we show in the online appendix that our qualitative empirical results also hold when targeting GDP growth at the quarterly frequency. Third, given that these papers have already established the forecasting benefits of high-dimensional QRs, our paper focuses on the economic insights that can be gained from ML methods.

Throughout, our ML method of choice is Quantile Partial Correlation Regression (QPCR) for two reasons. First, QPCR accommodates a large number of macrofinancial predictors while preserving the key QR advantages of interpretability and robustness to arbitrary distributions. Second, it offers theoretical guarantees--such as variable selection consistency--under time-series data, which provide additional reassurance when attaching interpretation to our empirical results \citep{JiHyungHongqi}. Moreover, we show that QPCR performs competitively against other ML methods, including non-linear models such as random forests, as well as volatility models.


Our empirical findings can be summarised as follows. We confirm the importance of financial indicators as predictors of downside growth risk, even after accounting for a comprehensive set of non-financial macroeconomic variables. We also identify additional drivers of growth vulnerabilities—namely, capacity utilisation, labour market conditions, and the housing market. Furthermore, the interpretability of QPCR enables us to validate and, importantly, quantify commonly held views regarding key predictors of downside growth risks, including the role of labour-market slack, yield curve dynamics, and credit spreads. In the online appendix, we also show that financial conditions provide information about growth tailwinds, although the relevant predictors differ from those associated with downside risks.

The second contribution of this paper is to decompose growth vulnerabilities into their individual components, and provide an easily interpretable, trackable, and comparable summary of our ML analysis in the form of sector-specific indices. Since our indices--and our FCI in particular--target a given quantile of future IP growth using a variable-selection method, they complement the targeted principal-components-based FCI proposed in \citet{AdrianDuarte,Sokol} as well as the Goldman Sachs FCI. However, our FCI offers two distinct advantages. First, it does not suffer from look-ahead bias, since it is not estimated using the entire sample but rather in a pseudo-out-of-sample fashion. Second, our index also partly addresses the concern raised in \citet{plagborg2020growth} that FCIs are endogenous to broader macroeconomic developments, and hence may contain non-financial information. We achieve this by exploiting the linear structure of QPCR: incorporating a broad set of variables ensures that the coefficients on financial variables capture their marginal effects net of the information in other variables. We note that accounting for the information contained in other sectors is especially important when creating an FCI targeted to predict a specific quantile of future growth. Indeed, the concern that an FCI may be primarily capturing information related to non-financial variables is heightened when it is constructed by targeting particular quantiles of a non-financial variable such as IP growth. Our FCI hence also bears some similarity to the Adjusted NFCI, which controls for non-financial macroeconomic variables in a similar way, but does not target specific quantiles \citep{ANFCI}. 



Our financial, labour‐market, and housing indices are strongly correlated with established benchmarks in these respective sectors, while being at most weakly correlated with indicators from other sectors, which enables them to isolate each sector’s predictive information. By contrast, the NFCI shows significant correlations with standard labour‐market and housing measures, which suggests that it also incorporates non-financial signals. Taken together, these findings highlight the value of constructing dedicated, sector‐specific indices using our proposed approach.

The rest of this paper is organised as follows. Section \ref{Section-Methodology} provides an overview of QPCR and the other approaches we use for comparison. Section \ref{Section-Simulation} provides simulation evidence in favour of evaluating GaR at the monthly frequency using IP growth. Section \ref{Section-Empirics} presents the results. The last section concludes.

%% file: sections/methodology_monthly.tex
\section{Model and methodology \label{Section-Methodology}}


We study the predictive relationship between the scalar sequence of IP Growth rates, $\{Y_{t+1}\}_{t = 1}^T$, and a sequence of stationary $\beta$-mixing predictors $\left\{ X_{t}\right\} _{t=1}^{T}$, where $X_{t}=\left(X_{t,1},\dots ,X_{t,p}\right)^{'}\in \mathbb{R}^{p}$, and $p$ can be larger than $T$. Throughout, we assume that $\{Y_{t+1}, X_{t}\}_{t = 1}^T$ and $X_{T+1}$ are observed.


For all models except for the GARCH model discussed below, we select covariates to predict the conditional quantile $\tau\in (0, 1)$ of $Y_{t+1}$ given $X_{t}$, which we model as
\begin{equation*}
Q_{Y_{t+1}}\left( \tau |X_{t}\right) =X_{t}^{'}\beta _{\tau }.
\end{equation*}%
For simplicity, we suppress the \( \tau \) subscript in \( \beta_{\tau} \), and use \( j \) instead of \( t \) in \( X_{j} \) to represent a particular covariate, denoted as \( \left( X_{1,j}, \dots, X_{T,j} \right)^{\prime} \in \mathbb{R}^{T} \) for \( j = 1, \dots, p \). Following standard notation, the quantile loss function is defined as 
$\rho_{\tau}(u) = u \left( \tau - 1\left( u < 0 \right) \right)$,
and its subgradient is given by 
$\psi_{\tau}(u) = \tau - 1\left( u < 0 \right).$

\subsection{Quantile Partial Correlation Regression}

In order to outline QPCR as used in \citet{ma2017variable, JiHyungHongqi}, we introduce the following notation and definitions. Let $S\subseteq \{1,\dots ,p\}$ denote a generic
index set of the covariate vector $X_{t}$, and $X_{t,S}$ denote the elements of $X_t$ indexed by $S$. We define the population quantile partial correlation (QPC) at the $\tau $-th quantile level for random variables $\left\{
Y_{t+1},X_{t,j},X_{t,S}\right\} $ as: 
\begin{align*}
qpcor_{\tau }\left( Y_{t+1},X_{t,j}|X_{t,S}\right) & =\frac{cov\left(
\psi _{\tau }\left( Y_{t+1}-X_{t,S}^{'}\alpha _{S}^{0}\right)
,X_{t,j}-X_{t,S}^{'}\theta _{j, S}^{0}\right) }{\sqrt{var\left( \psi _{\tau
}\left( Y_{t+1}-X_{t,S}^{'}\alpha_{S}^{0}\right) var\left(
X_{t,j}-X_{t,S}^{'}\theta _{j, S}^{0}\right) \right) }} \\
& =\frac{E\left[ \psi _{\tau }\left( Y_{t+1}-X_{t,S}^{'}\alpha
_{S}^{0}\right) \left( X_{t,j}-X_{t,S}^{'}\theta _{j, S}^{0}\right) \right] 
}{\sqrt{\tau \left( 1-\tau \right) \sigma _{t,j}^{2}}},
\end{align*}%
where $\alpha _{S}^{0}=\arg \min_{\alpha _{S}}E\left( \rho _{\tau }\left(
Y_{t+1}-X_{t,S}^{'}\alpha_{S}\right) \right) $, $\theta _{j, S}^{0}=\arg
\min_{\theta _{j, S}}E\left( \left( X_{t,j}-X_{t,S}^{'}\theta
_{j, S}\right) ^{2}\right) $, and $\sigma _{t,j}^{2}=var\left( X_{t,j}-X_{t,S}^{'}\theta
_{j, S}\right) $. Its sample analogue, denoted $\widehat{qpcor}_{\tau }$, is given by
\begin{equation*}
\widehat{qpcor}_{\tau }\left( Y_{t+1},X_{t,j}|X_{t,S}\right) =\frac{%
\frac{1}{T}\sum_{t=1}^{T}\left( \psi _{\tau }\left(
Y_{t+1}-X_{t,S}^{'}\hat{\alpha}_{S}\right) \left(
X_{t,j}-X_{t,S}^{'}\hat{\theta}_{j, S}\right) \right) }{\sqrt{\tau
\left( 1-\tau \right) \hat{\sigma}_{t,j}^{2}}},
\end{equation*}%
where $\hat{\alpha}_{S}=\arg \min_{\alpha _{S}} \frac{1}{T}\sum_{t=1}^{T}\rho
_{\tau }\left( Y_{t+1}-X_{t,S}^{'}\alpha _{S}\right) $ is the QR estimator of the quantile regression of $Y_{t+1}$ on $X_{t,S}$, $\hat{%
\theta}_{j, S}=\arg \min_{\theta_{j,S}} \frac{1}{T}\sum_{t=1}^{T}\left(
X_{t,j}-X_{t,S}^{'}\theta _{j, S}\right) ^{2}$ is the OLS estimator of the projection of $X_{t,j}$ on $X_{t,S}$, and $\hat{\sigma}%
_{t,j}^{2}=\frac{1}{T}\sum_{t=1}^{T}\left( X_{t,j}-X_{t,S}^{'}\hat{%
\theta}_{j, S}\right) ^{2}$ is the variance estimator of the residual in the projection of $X_{t,j}$ on $X_{t,S}$. Finally, let  $\hat{\varrho}_{j,k}$ be the sample correlation coefficient between $X_j$ and $X_k$, and for every $j = 1,\dots, p$, define $\hat{j}(m)\neq j$ as the index of the variable such that $|\hat{\varrho}_{j,\hat{j}(m)}|$ is the $m$-th largest value in the set $\{|\hat{\varrho}_{j, l}|\}_{l = 1, l\neq j}^p$, as well as the confounding set
\begin{align*}
S_j^{\nu}(m) :=\; & S^\nu_j\left(\{X_t\}_{t=1}^T, m\right) \\
=\, & \left\{ k \in \{1, \dots, p\} : k \neq j \text{ and } 
\left| \hat{\varrho}_{j,k} \right| \geq |\hat{\varrho}_{j,\hat{j}(m)}| \right\}.
\end{align*}
With this notation and definitions in place, Algorithm \ref{algo:QPCR} describes QPCR. Intuitively, the algorithm iteratively selects the variables that are most predictive of the $\tau$-th quantile of the dependent variable, conditional on the previously selected variable as well as the variables most highly correlated with it. We note that QPCR offers a series of advantages over competing ML methods. First, as shown in \citet{JiHyungHongqi} under suitable assumptions that allow for time series, QPCR is theoretically guaranteed to eventually select all relevant predictors even if the number of predictors outstrips the number of observations ($p > T$). Second, QPCR helps mitigate the problem of correlated predictors that can affect many popular ML methods such as the LASSO (see, e.g., \citet{sun2024sorted}), since the inclusion of the conditioning set ensures that candidate predictors that are highly correlated with previously selected predictors are not selected.


The QPCR algorithm depends on hyperparmeters, which we set as follows. The confounding set is updated $d^{\ast }$ $=\left\lfloor{\frac{T}{\log 
{T}}}\right \rfloor ^{\frac{1}{2}}$ times, and, following \citet{ma2017variable}, the 
total number of iterations is $D_{max}=\left\lfloor \frac{T}{\log {T}}\right\rfloor $. The size of the confounding set is controlled by the parameters $m_d$ for $d = 1, \dots, d^*$, which we set equal to $m_{d} =  {\left\lfloor \frac{T}{\log {T}}
\right \rfloor }^{\frac{1}{2}}$. We note that our choice of $D^*$ in Algorithm \ref{algo:QPCR}, which determines the number of predictors, corresponds to choosing the number of predictors that minimises the extended Bayesian information criterion (EBIC), and we choose $C = 1$.

\begin{algorithm}[H]
\caption{Quantile Partial Correlation Regression}\label{algo:QPCR}
\begin{algorithmic}[1]
\State Choose the quantile level $\tau \in (0, 1)$, the number of updates to the confounding set $d^*\in \mathbb{N}$, the maximum number of predictors to be added to the confounding set at each update $\{m_j\in\mathbb{N}\}_{j = 1}^{d^*}$, the maximum number of iterations $D_{max}\in \mathbb{N}$, and the EBIC constant $C\in \mathbb{R}$
\State Initialize active set $S_1 = \emptyset$ and $S_1^{\nu}(m_1) = \emptyset$
\For{$d = 1$ to $D_{\text{max}}$}
    \State Set $\bar{S}_d = S_d \cup S_{\tilde d}^{\nu}(m_{\tilde d})$ for $\tilde d = \text{min}(d, d^*)$
    \State Select covariate index
    \[
    j^{\ast} = \arg\max_{j \notin \bar{S}_d} \left| \widehat{qpcor}_\tau(Y_{t+1}, X_{t,j} \mid X_{t,\bar{S}_d}) \right|
    \]
    \State Update $S_{d+1} = S_d \cup \{ j^{\ast} \}$
\EndFor
\State Calculate
\[
D^* = \underset{D\in \mathbb{N}, D \leq D_{\text{max}}}{\text{arg min}} \log\left(\frac{1}{T}\sum_{t=1}^{T}\rho_\tau\left(Y_{t}-X_{t, S_D}^{'}\hat{\beta}_{S_D}^{QPCR}\right)\right)+C \frac{\log T \log D}{T},
\]
and 
\[
\hat{\beta}_{S_D}^{QPCR} =\underset{\beta_{S_D}\in \mathbb{R}^{|S_D|}}   {\text{arg min}} \frac{1}{T} \sum_{t=1}^T \rho_\tau(Y_{t+1} - X_{t,S_D}' \beta_{S_D})
\]
\State Calculate the estimated Quantile Partial Correlation Regression coefficient as the vector $\hat{\beta}^{QPCR} \in \mathbb{R}^p$ where $\hat{\beta}^{QPCR}_j = \hat{\beta}_{S_{D^*}, j}^{QPCR}$ if $j \in S_{D^*}$ and $\hat{\beta}^{QPCR}_j  = 0$ otherwise
\State Calculate the conditional quantiles $Q_{Y_{t+1}}^{QPCR}\left( \tau |X_{t}\right) = X_t'\hat{\beta}^{QPCR}$ for $t = 1, \dots, T + 1$.
\end{algorithmic}
\end{algorithm}

\subsection{Other methods}

This section presents additional methods used in our forecasting exercise, including state-of-the-art quantile ML forecasting techniques and standard volatility models. We consider three linear penalization-based quantile ML methods: the $l_{1}$-penalized LASSO ($l_1$-QR) following \citet{belloni2011,tibshirani1996regression}, the smoothly-clipped absolute deviation (SCAD) penalty from \citet{fan2001variable}, and the minimax concave penalty (MCP) proposed by \citet{zhang2010nearly}. While these techniques lack theoretical guarantees for time series data, they are widely applied in high-dimensional regression models. We also include Quantile Random Forests (QRFs), a nonlinear ML approach that outperforms Quantile Neural Nets in our forecasting exercises,\footnote{In particular, using QNNs with 3 to 24 hidden nodes and ReLU activation, we found that the forecasting performance of QNNs is worse than that of QPCR and QRFs. Moreover, QNNs suffer from several limitations, including a lack of interpretability, and a large number of hyperparameters that require tuning. For these reasons, we do not include QNNs in the main text.} and to the best of our knowledge has not yet been considered in the GaR literature (see \citet{lenza2023density} for an application to inflation forecasting). Finally, we consider a standard GARCH model as employed in \citet{brownlees2021backtesting}.


\subsubsection{Penalized quantile regression}
For all three penalized QR methods, the estimator takes the form
\begin{equation*}
\hat{\beta }^{i}=\underset{\beta }{\arg \min }\frac{1}{%
T}\sum_{t=1}^{T}\rho _{\tau }\left( Y_{t+1}-X_{t}^{'}\beta %
\right) +\sum_{j=1}^{p}q_{\lambda, a}\left( \beta _{j}\right),
\end{equation*}%
and the conditional quantile takes the form $Q_{Y_{t+1}}^{i}\left( \tau |X_{t}\right) = X_t'\hat{\beta}^{i}$ for $t = 1, \dots, T + 1$,
where $q_{\lambda, a}\left( \beta _{j}\right)$ is a penalty function depending on scalars $\lambda \geq 0$ and $a\geq 0$, and $i \in \{l_1\text{-QR, SCAD, MCP}\}$. 

For $l_1$-QR, the penalty function is $$q_{\lambda, a}\left( \beta _{j}\right) =\lambda \left\vert \beta
_{j}\right\vert.$$ The SCAD method employs the penalty function $q_{\lambda, a}(\beta)$, whose derivative is given by:
\begin{equation*}
q^{'}_{\lambda, a}\left( \beta \right) =\lambda \left[ \boldsymbol{1}\left( \beta
\leq \lambda \right) +\frac{ \max \left\{ a\lambda -\beta , 0 \right\} } {\left(
a-1\right) \lambda }\boldsymbol{1}\left( \beta >\lambda \right) \right] ,
\end{equation*}
where $a>2$ is a tuning parameter controlling the nonconvexity of the penalty. The MCP method employs the penalty function $q_{\lambda, a}(\beta)$, with its derivative defined as:
\begin{equation*}
q^{'}_{\lambda, a}\left( \beta \right) =%
\begin{cases}
\begin{array}{c}
sgn\left( \beta \right) \left( \lambda -\frac{\left\vert \beta \right\vert }{%
a}\right)  \\ 
0%
\end{array}
& 
\begin{array}{c}
if\;\left\vert \beta \right\vert \leq a\lambda  \\ 
otherwise%
\end{array}%
\end{cases}%
\end{equation*}%
where $a>1$ is a tuning parameter that controls the decay speed of the penalty. Following standard practice in the literature, we use the 5-fold cross-validation criterion to choose the optimal $\lambda $ and $a$. We note that contrarily to QPCR, these methods do not come with any theoretical guarantees under time series.

\subsubsection{Quantile random forests}

In this subsection, we briefly review the QRF methods proposed by \citet{meinshausen2006quantile} and \citet{athey2019generalized}, which we term QRFM and QRFATW, respectively. Algorithm 2 provides the implementation details. Intuitively, the key idea is to construct an ensemble of decision trees that assign weights to past observations based on their similarity to the vector of covariates used in the prediction step, and then estimate the conditional quantile via a weighted quantile loss function. QRFM and QRFATW follow a similar tree-based structure, but differ in their splitting criteria: QRFM seeks to maximise the difference in conditional means between child nodes, while QRFATW maximises the differences in empirical conditional quantiles.

We set the number of trees to $B = 2000$ and use a minimum terminal node size of $m = 5$ as the stopping criterion. At each split, a random subset of $\lfloor p^\frac{1}{2} \rfloor + 20$ covariates is considered for selecting the splits (i.e., $J = 20$).

\noindent

\rule{\linewidth}{1pt}
\vskip -2em
\textbf{Algorithm 2:} \textit{Quantile Random Forests (QRFM and QRFATW)} \label{algo:QRF}\\
\vskip -4.8em
\rule{\linewidth}{0.6pt}

\algblock{Using}{EndUsing}

\begin{algorithmic}[1]

\State Choose the quantile level $\tau \in (0, 1)$, the number of trees $B\in \mathbb{N}$, minimum leaf size $m\in \mathbb{N}$, the constant to be added to feature subsample size $J\in \mathbb{N}$, and the conditioning predictors $x\in \mathbb{R}^p$
\For{$b = 1$ to $B$}
    \State Draw a sample $\mathcal{D}_b =\{(Y_{t+1}^*, X_t^*)\}_{t = 1}^{\lfloor T/2\rfloor}$ without replacement from $\{(Y_{t+1}, X_t)\}_{t = 1}^T$
    \State Randomly split $\mathcal{D}_b$ into two samples of equal size, $\mathcal{D}_b^{\text{tree}}$ and $\mathcal{D}_b^{\text{weights}}$

    \State Initialize queue: $\texttt{Queue} \gets \{(\mathcal{D}_b^{\text{tree}}, \texttt{Path} = \emptyset)\}$, $k = 0$

    \While{\texttt{Queue} is not empty}
        \State Pop $(\mathcal{D}_{b, \text{leaf}}, \texttt{Path})$ from \texttt{Queue}\footnote{That is, take the last element from \texttt{Queue}, $(\mathcal{D}_{b, \text{leaf}}, \texttt{Path})$, and remove it from \texttt{Queue}}
        \If{$|\mathcal{D}_{b, \text{leaf}}| \leq m$}
            \State Store $\mathcal{W}_{b,k} := \left\{ x \in \mathbb{R}^p : \text{$x$ satisfies all conditions in ordered } \texttt{Path} \right\}$
            \State Set $k = k + 1$
        \Else
            \State Select $\mathcal{S} \subseteq \{1, \dots, p\}$ uniformly at random with $|\mathcal{S}| = \lfloor \sqrt{p} \rfloor + J$
            \State For each $j \in \mathcal{S}$ and split value $s$, define:
            \begin{align*}
                \mathcal{L}_{j,s} &= \{(Y_{t+1}^*, X_t^*) \in \mathcal{D}_{b,\text{leaf}} : X_{t,j}^* \leq s\} \\
                \mathcal{R}_{j,s} &= \{(Y_{t+1}^*, X_t^*) \in \mathcal{D}_{b,\text{leaf}} : X_{t,j}^* > s\}
            \end{align*}

            \If{QRFM (mean-based split)}
                \begin{align*}
                    j^*, s^* &= \arg\max_{j \in \mathcal{S}, s \in \mathbb{R}} \sum_{\mathcal{C} \in \{\mathcal{L}_{j,s}, \mathcal{R}_{j,s}\}} |\hat{Y}(\mathcal{C}) - \hat{Y}(\mathcal{D}_{b,\text{leaf}})| \\
                    \hat{Y}(\mathcal{K}) &= \frac{1}{|\mathcal{K}|} \sum_{(Y_{t+1}^*, X_t^*) \in \mathcal{K}} Y_{t+1}^* \quad \text{for} \quad \mathcal{K}\in\{\mathcal{L}_{j, s},  \mathcal{R}_{j, s}, \mathcal{D}_{b,\text{leaf}}\}
                \end{align*}
            \ElsIf{QRFATW (quantile-based split)}
                \begin{align*}
                    j^*, s^* &= \arg\max_{j \in \mathcal{S}, s \in \mathbb{R}} \sum_{\mathcal{C} \in \{\mathcal{L}_{j,s}, \mathcal{R}_{j,s}\}} |\hat{\mathcal{Q}}(\mathcal{C}, \tau) - \hat{\mathcal{Q}}(\mathcal{D}_{b,\text{leaf}}, \tau)| \\
                    \hat{\mathcal{Q}}(\mathcal{K}, \tau) &= \inf \left\{ y \in \mathbb{R}: \frac{1}{|\mathcal{K}|} \sum_{(Y_{t+1}^*, X_t^*) \in \mathcal{K}} \mathbf{1}(Y_{t+1}^* \leq y) \geq \tau \right\}\\
                    & \quad \quad  \text{for} \quad \mathcal{K}\in\{\mathcal{L}_{j, s},  \mathcal{R}_{j, s}, \mathcal{D}_{b,\text{leaf}}\}
                \end{align*}
            \EndIf
            \State Append $(\mathcal{L}_{j^*, s^*}, \texttt{Path} \gets \texttt{Path appended with } (j^*, s^*, \texttt{left})$ to \texttt{Queue}
            \State Append $(\mathcal{R}_{j^*, s^*},\texttt{Path} \gets \texttt{Path appended with } (j^*, s^*, \texttt{right})$ to \texttt{Queue}
        \EndIf
    \EndWhile
    \State Save the number of leaves in tree $b$, $l_b = k$
\EndFor
\State Calculate $\{\mathcal{W}_{b,k(x)}\}_{b = 1}^B$ such that $x \in \mathcal{W}_{b,k(x)}$ for all $b = 1, \dots, B$

\State Compute weights
\[
w_t(x) = \frac{1}{B}\sum_{b = 1}^B\left|\left\{(Y_{t+1}^*, X_t^*)\in\mathcal{D}_b^{\text{weights}}: X_t^*\in  \mathcal{W}_{b,k(x)}\right\} \right|
\]
\State Return
\[
\hat{Q}_{Y_{t+1}}\left( \tau \mid x \right)
= \underset{q\in \mathbb{R}}{\arg\min}\left\Vert
\sum_{t=1}^{T}w_{t}\left(x\right)\left(\tau-\boldsymbol{1}%
\left(Y_{t+1}\leq q \right)\right)\right\Vert _{2} 
\]

\end{algorithmic}
\vskip -2.1em
\rule{\linewidth}{0.6pt}

\pagebreak

\subsubsection{GARCH}

In addition to ML quantile regression methods, we also consider a standard GARCH approach as in \citet{brownlees2021backtesting},
\begin{equation*}
Y_{t+1}=\mu _{t+1|t}+\sigma _{t+1|t}Z_{t+1},
\end{equation*}%
where $Z_{t+1} \overset{i.i.d.}{\sim} \mathcal{D}(0,1)$, and $\mathcal{D}(0,1)$ is a location-scale distribution with mean 0 and variance 1. $\mu _{t+1|t}$ and $\sigma _{t+1|t}$ are the mean and standard deviation of $Y_{t+1}$ conditional on information available at time $t$. The conditional quantile can then be expressed as
\begin{equation*}
{Q}_{Y_{t+1}}\left( \tau |\mathcal{F}_{t}\right) =\mu _{t+1|t}+\sigma
_{t+1|t}F_{\mathcal{D}}^{-1}(\tau), 
\end{equation*}%
where $ F_{\mathcal{D}}^{-1}(\cdot)$ is the inverse cumulative density function of the distribution $\mathcal{D}(0,1)$, and $\left\{ \mathcal{F}_{t}\right\} _{t \geq 0}$ is the natural filtration (information set) at time $t$. We follow the bootstrapped simulation approach suggested in \citet[Algorithm 1]{brownlees2021backtesting} to determine  $ F_{\mathcal{D}}^{-1}(\tau)$. We formulate an AR(1) model for the conditional mean, $\mu _{t+1|t}=\phi _{0}+\phi _{1}Y_{t}$, and a GARCH(1,1) model for the conditional variance, $\sigma _{t+1|t}^{2}=\omega +\alpha \left(Y_{t}-\mu _{t|t-1}\right) ^{2}+\gamma \sigma _{t|t-1}^{2}$.

%% file: sections/simulation.tex
\section{Simulations \label{Section-Simulation}}


We conduct stylised simulations to show that large samples are essential for correctly identifying relevant predictors in tail quantiles, justifying our focus on assessing growth vulnerabilities at the monthly frequency using IP Growth. To highlight that this issue does not stem from an especially unfavourable data-generating process (DGP), we employ a very simple DGP that is not necessarily intended to represent GaR dynamics.


We generate predictors  as the absolute values of independent draws from a multivariate Normal distribution to ensure non-negativity in the location-scale linear quantile regression formulation, and, for simplicity, we set $X_{t, j} \overset{i.i.d.}{\sim}|\mathcal{N}(0, 1)|$ for $t = 1, \dots, T$ and $j = 1, \dots p$. We model the response variable \(y_{t+1}\) by a location-scale framework
\[
Y_{t+1} = X_t'\alpha_{t} + \left( X_t'\beta_t \right)\epsilon_t,
\]
where \(\alpha_{t}\) and \(\beta_{t}\) are $p$-dimensional column vectors with entries $\alpha_{t, j},\beta_{t, j}$, and $\epsilon_t\overset{i.i.d.}{\sim}\mathcal{N}(0, 1)$. Consequently, the conditional linear quantile function for \(Y_{t + 1}\) is
\[
Q_{Y_{t+1}}(\tau|X_t)  = X_t'(\alpha_t +  \beta_t \Phi^{-1}(\tau)).
\]
We consider a fixed sparse, a time-varying sparse, and a dense setup. Under the fixed sparse setup, for all $t = 1, \dots, T$, $\beta_{t, j} = 1$, $\alpha_{t, j} = -1$ for $j = 1, \dots, s$, and $\beta_{t, j} = \alpha_{t, j} = 0$ otherwise. For the time-varying sparse setup, $\beta_{t, j} = 1$, $\alpha_{t, j} = -1$ for $j = 1, \dots, s$, $t = 1, \dots, \lfloor T/2\rfloor$,  $\beta_{t, j} = 0.5$, $\alpha_{t, j} = -1$ for $j = 1, \dots, s$, $t = \lfloor T/2\rfloor + 1, \dots, T$, and $\beta_{t, j} = \alpha_{t, j} = 0$ otherwise. For the dense setup, for all $t = 1, \dots, T$, $\beta_{t, j} = 1$, $\alpha_{t, j} = -1$ for $j = 1, \dots, s$, and $\beta_{t, j} = \alpha_{t, j} = 1/p$ otherwise. For all DGPs, we denote the predictors indexed by $j = 1, \dots, s$ as the `relevant predictors'. Thus, all setups feature $s$ relevant predictors, but differ in terms of the informativeness of the other predictors, as well as the stability of the coefficients over time.

For all setups, we set  $s = 5$ and \(\tau = 0.05\). To illustrate that large samples are needed to correctly identify relevant predictors in tail quantiles, we consider values of $T$ and $p$ that roughly match the FRED database at the monthly and quarterly frequency, i.e., $T = 500, p = 110$ and $T = 100, p = 220$, respectively. For each of the selection-based ML methods presented in Section \ref{Section-Methodology}, we report the selection frequency of the relevant predictors, as well as the average number of times non-relevant predictors are selected (`false selections') across 1000 simulations. 

Table \ref{Table-Sims} presents the simulation results. For $T = 500, p = 110$, all selection-based methods select the relevant predictors with high frequency, and QPCR has the lowest average false selection rate. In contrast, for $T = 100, p = 220$, all selection-based methods fail to reliably select the relevant predictors, and the average false selection rate is similarly high across all methods. Motivated by these findings, our main empirical analysis focuses on monthly IP‐growth (420 observations) rather than quarterly GDP‐growth (only 111 observations). Nevertheless, in the online appendix, we also report our estimation results using GDP growth as the dependent variable.

\begin{table}
\begin{centering}
\par\end{centering}
\begin{centering}
\caption{Simulation results}
\label{Table-Sims}
\par\end{centering}
\scriptsize
\renewcommand{\arraystretch}{1.3}
\begin{centering}
\begin{tabular}{cc|ccccccccccccc}
\hline 
\multirow{2}{*}{Setup} & \multirow{2}{*}{Method} 
  & \multicolumn{6}{c}{$T=500,p=110$} 
  & \multicolumn{7}{c}{$T=100,p=220$}\\
 &  & $X_{1}$ & $X_{2}$ & $X_{3}$ & $X_{4}$ & $X_{5}$ & 
   \shortstack{Avg.\\ false}  
  &  & $X_{1}$ & $X_{2}$ & $X_{3}$ & $X_{4}$ & $X_{5}$ & 
   \shortstack{Avg.\\ false}\\
\cline{1-8} \cline{10-15}
\multirow{4}{*}{Fixed sparse} 
  & QPCR      & 0.90 & 0.88 & 0.90 & 0.90 & 0.89 &  1.17 
              &  & 0.10 & 0.10 & 0.10 & 0.10 & 0.12 & 3.47\\
  & $l_{1}$-QR& 0.97 & 0.97 & 0.97 & 0.97 & 0.97 & 15.12 
              &  & 0.14 & 0.12 & 0.14 & 0.12 & 0.15 & 2.65\\
  & SCAD      & 0.95 & 0.95 & 0.95 & 0.94 & 0.95 &  7.32 
              &  & 0.14 & 0.13 & 0.14 & 0.14 & 0.15 & 2.97\\
  & MCP       & 0.95 & 0.95 & 0.95 & 0.94 & 0.95 &  7.54 
              &  & 0.15 & 0.15 & 0.16 & 0.13 & 0.16 & 3.09\\
\cline{1-8} \cline{10-15}
\multirow{4}{*}{\shortstack{Time-varying\\sparse}} 
  & QPCR      & 0.88 & 0.86 & 0.86 & 0.86 & 0.86 &  1.42 
              &  & 0.10 & 0.12 & 0.10 & 0.12 & 0.11 & 3.61\\
  & $l_{1}$-QR& 0.94 & 0.94 & 0.94 & 0.94 & 0.94 & 14.57 
              &  & 0.14 & 0.13 & 0.14 & 0.12 & 0.15 & 2.92\\
  & SCAD      & 0.93 & 0.92 & 0.91 & 0.92 & 0.92 &  7.57 
              &  & 0.14 & 0.13 & 0.14 & 0.14 & 0.16 & 3.05\\
  & MCP       & 0.92 & 0.92 & 0.91 & 0.91 & 0.92 &  7.87 
              &  & 0.14 & 0.15 & 0.15 & 0.13 & 0.16 & 3.20\\
\cline{1-8} \cline{10-15}
\multirow{4}{*}{Dense} 
  & QPCR      & 0.81 & 0.81 & 0.81 & 0.80 & 0.79 &  2.17 
              &  & 0.09 & 0.08 & 0.07 & 0.09 & 0.08 & 3.61\\
  & $l_{1}$-QR& 0.91 & 0.93 & 0.92 & 0.91 & 0.91 & 13.95 
              &  & 0.12 & 0.11 & 0.11 & 0.10 & 0.13 & 2.65\\
  & SCAD      & 0.86 & 0.87 & 0.86 & 0.85 & 0.86 &  7.29 
              &  & 0.13 & 0.11 & 0.13 & 0.11 & 0.14 & 2.96\\
  & MCP       & 0.86 & 0.86 & 0.84 & 0.83 & 0.86 &  7.49 
              &  & 0.12 & 0.11 & 0.12 & 0.10 & 0.14 & 2.95\\
\hline 
\end{tabular}

\par\end{centering}
\centering{}

\begin{note*}
\footnotesize
\textit{Selection frequencies for the five relevant predictors ($X_1$ to $X_5$) and average false selections for different selection-based ML methods and simulation setups over 1000 simulations. QPCR denotes to the quantile partial correlation regression. $l_1$-QR, SCAD, and MCP refer to penalized quantile regressions using the $l_1$ penalty, smooth-clipped absolute deviation penalty, and minimax concave penalty, respectively.}
\end{note*}
\end{table}

%% file: sections/empirics_monthly.tex
\section{Empirical results \label{Section-Empirics}}

Throughout this section, we consider a rolling-window pseudo-out-of-sample forecasting exercise for $\tau = 0.05$ on monthly data from the Fred-MD. Our sample spans January 1971 through October 2024, and we transform the data as recommended in \citet{mccracken2020fred}. Each estimation window contains 420 observations and 111 predictors. Beginning in January 2006, we produce one-step-ahead forecasts for each month up to October 2024, yielding 225 out-of-sample predictions. In Section \ref{Section-Horse}, we compare the forecasting performance of QPCR to the other methods described in Section \ref{Section-Methodology}, showing that QPCR performs competitively. In Section \ref{Section-Results} we analyse the drivers of downside risk over time. In Section \ref{section-GaR-Decomp} we decompose growth vulnerabilities into their individual components, and construct sector-specific indices. The online appendix contains analogous results for upside risks, i.e., $\tau = 0.95$.

\subsection{Forecasting performance \label{Section-Horse}}


We evaluate the forecasting performance using the mean quantile prediction error (MPE) defined as
\begin{equation*}
   MPE = \frac{1}{T_{out}}\sum_{t=1}^{T_{out}} \rho_\tau (y_{t}-\hat{y}_{t,\tau}),
\end{equation*}
where $T_{out}$ is the total number of out-of-sample forecasting period, $y_{t}$ is the IP growth rate one period after the last observation in the estimation sample, and $\hat{y}_{t,\tau}$ is its predicted conditional $\tau$-quantile. We also evaluate standard Diebold-Mariano tests comparing QPCR to the other methods outlined in Section \ref{Section-Methodology}, although we note that they may not have the usual interpretation when ML methods lack necessary theoretical guarantees.

The MPEs and DM statistics reported in Table \ref{Table-Forecasting} show that QPCR performs favorably compared to other ML methods, including non-linear QRF. We hence conclude that QPCR is highly suited to analysing GaR, since it comes with theoretical guarantees under time-series data, provides interpretable results by selecting relevant predictors, and performs competitively against other linear and non-linear ML and volatility models.

\begin{table}[H]

\centering
\caption{Forecasting performance, $\tau = 0.05$} 
\label{Table-Forecasting}
{\scriptsize
\begin{tabular}{c|c|c}
\hline
Method & MPE ($\times10^{-3}$) & DM-Statistics (QPCR/Others) \\
\hline
QPCR         & 1.462 & --       \\
$l_{1}$-QR   & 1.692 & -1.378   \\
SCAD         & 1.769 & -2.017   \\
MCP          & 1.759 & -1.855   \\
QRFATW       & 1.480 & -0.154   \\
QRFM         & 1.449 & 0.114    \\
GARCH        & 1.599 & -1.445   \\
\hline
\end{tabular}
}
\vspace{0.5em}
\footnotesize
\begin{note*}
\textit{Negative DM values indicate that QPCR outperforms the corresponding method. Under appropriate regularity conditions, DM test statistics are asymptotically normally distributed and the usual critical values of $\pm{1.96}$ apply. QPCR denotes to the quantile partial correlation regression. $l_1$-QR, SCAD, and MCP refer to penalized quantile regressions using the $l_1$ penalty, smooth-clipped absolute deviation penalty, and minimax concave penalty, respectively. QRFATW and QRFM represent the quantile random forests  proposed by \citet{athey2019generalized} and \citet{meinshausen2006quantile}, respectively.}
\end{note*}
\end{table}


\subsection{Drivers of downside risk \label{Section-Results}}

\begin{figure}[H]
    \centering
    \includegraphics[width=1\linewidth]{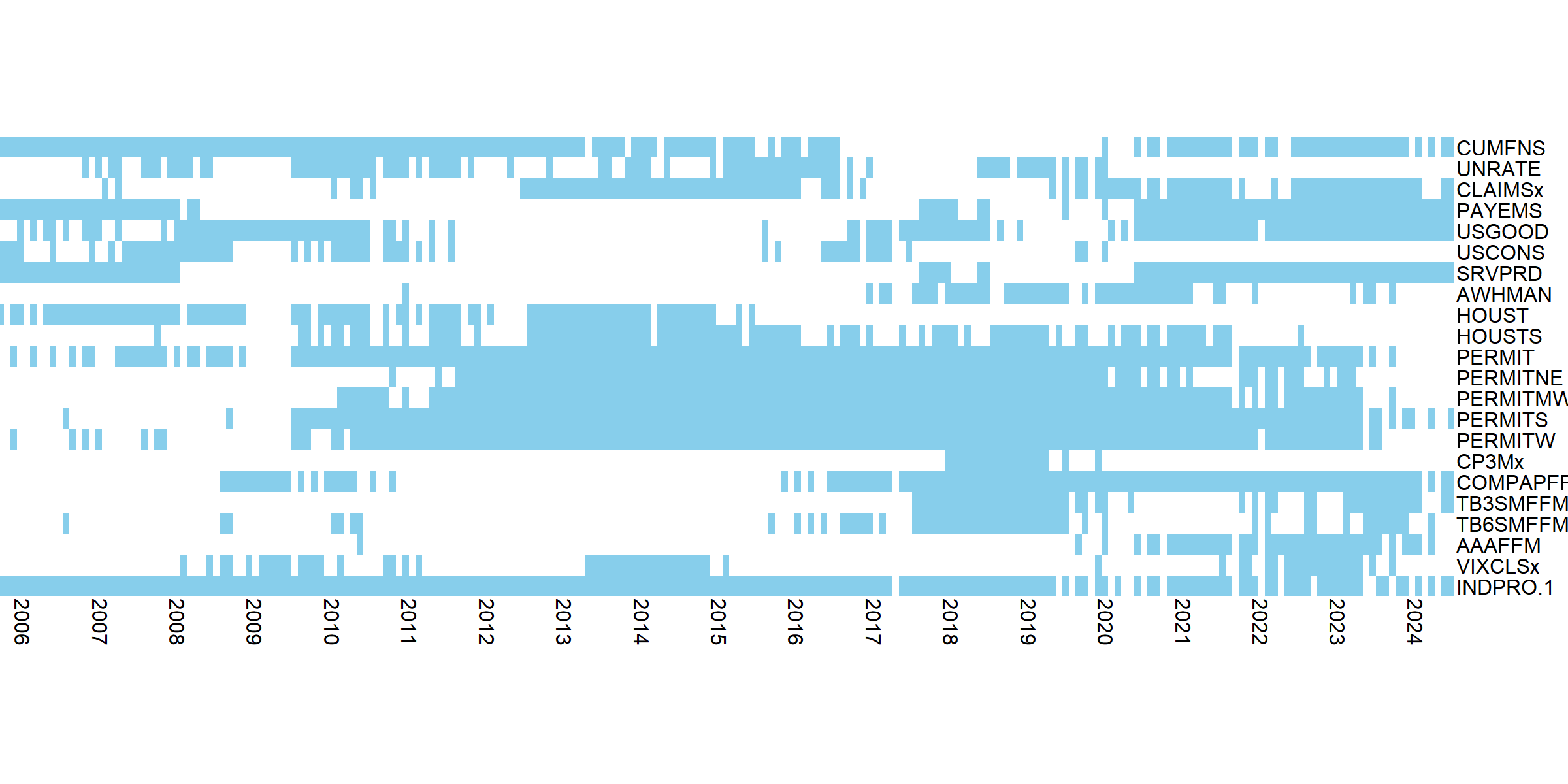}
    \caption{Selected variables for $\tau = 0.05$ }
    \label{Figure-QPCS-Select-Heat-0.05}
    \footnotesize
    \begin{note*}
        \textit{Variables selected for at least 12 consecutive months. Each row represents a selected variable, and blue cells indicate periods of selection. CUMFNS: capacity utilisation; UNRATE: unemployment rate; CLAIMSx: initial claims; PAYEMS: total nonfarm employees; USGOOD: goods-producing employees; USCONS: construction employees; SRVPRD: service-producing employees; AWHMAN: average weekly hours (manufacturing); HOUST: housing starts; HOUSTS: housing starts (south); PERMIT: new private housing permits; PERMITNE: new private housing permits (northeast); PERMITMW: new private housing permits (midwest); PERMITS: new private housing permits (south); CP3Mx: three-month AA financial commercial paper rate; COMPAPFFx: three-month commercial paper rate minus Federal Funds rate; TB3SMFFM: three-month Treasury rate minus Federal Funds rate; TB6SMFFM: six-month Treasury rate minus Federal Funds rate; AAAFFM: Moody's Aaa corporate bond rate minus Federal Funds rate; VIXCLs: VIX; INDPRO.1: lag of industrial production.}
    \end{note*}
\end{figure}

To avoid muddying our discussion by predictors that are only sporadically selected, we focus on those predictors that are selected in at least twelve consecutive months. The resulting heat map of these `systematically selected' predictors for $\tau = 0.05$ is shown in Figure \ref{Figure-QPCS-Select-Heat-0.05}. We find that the main drivers of downside risk (after controlling for the lag of industrial production) to GDP can be categorised into just four groups: capacity utilisation (CUMFNS), labour-market indicators (UNRATE, CLAIMSx, PAYEMS, USGOOD, SVPRD, AWHMAN), housing-related indicators (HOUST, HOUSTS, PERMIT), and financial indicators (CP3Mx, COMPAPFFx, TB3SMFFM, TB6SMFFM, AAFM, VIXCLSx). 

Figure \ref{Figure-QPCS-Select-Heat-0.05} also shows how the predictors of downside risk have changed over time. Housing and labour-market indicators are consistently selected. Financial variables exhibit time-varying importance, with notable spikes in selection during periods of financial stress such as the Global Financial Crisis (GFC) and the Covid-19 pandemic. For instance, the commercial paper spread and the VIX are selected consistently during and surrounding the great financial crisis. Moreover, at the end of the Covid-19 pandemic and during the subsequent inflation surge, spreads and the VIX are also consistently selected. The commercial paper spread (COMPAPFFx) is consistently selected starting in 2016, suggesting that changes in short-term financing conditions of companies have become an important transmission channel of monetary policy.

\begin{figure}[H]
    \centering
    \begin{tabular}{cc}
        \subfloat[CUMFNS]{
            \includegraphics[width=0.45\linewidth]{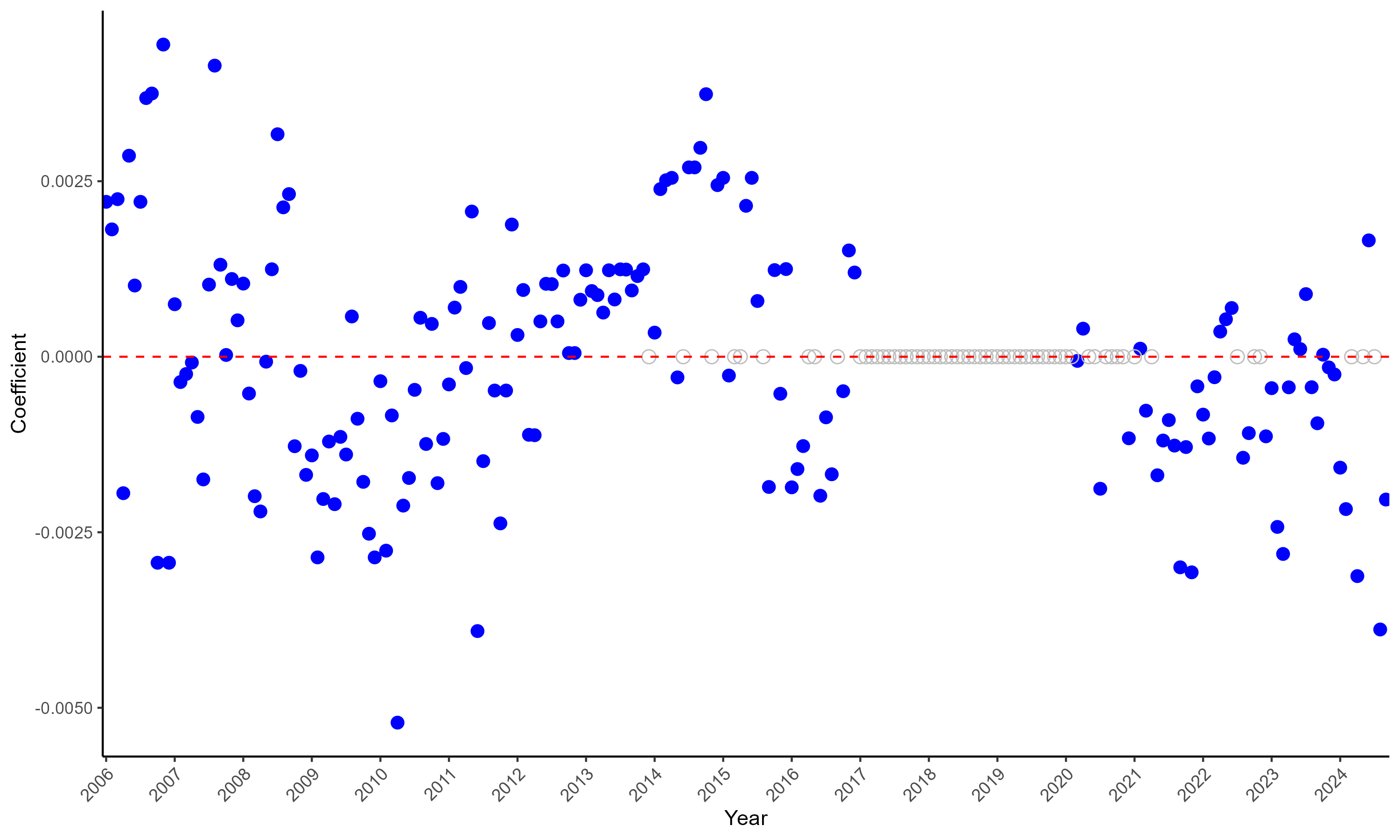}
        } &
        \subfloat[UNRATE]{
            \includegraphics[width=0.45\linewidth]{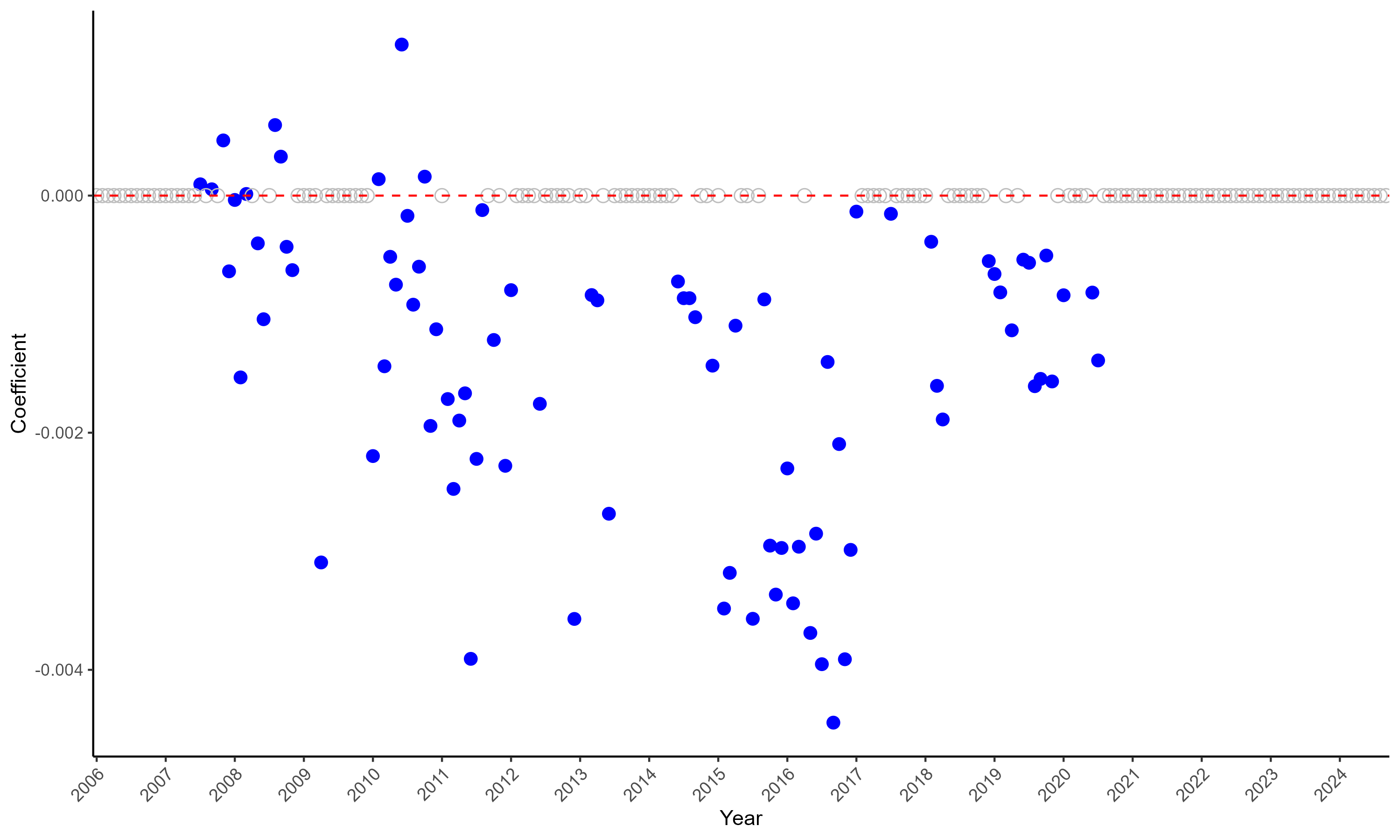}
        } \\
        \subfloat[CLAIMSx]{
            \includegraphics[width=0.45\linewidth]{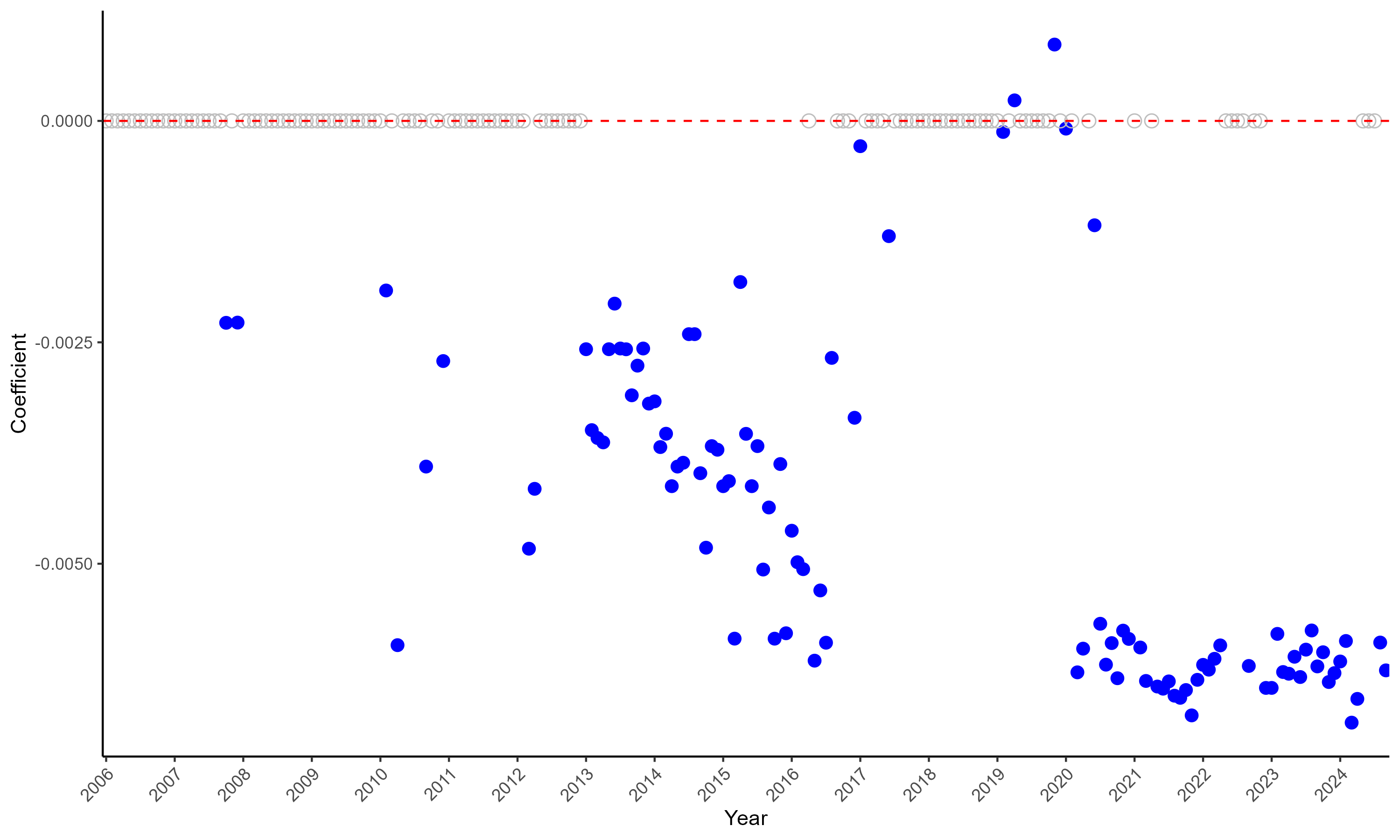}
        } &
        \subfloat[HOUST]{
            \includegraphics[width=0.45\linewidth]{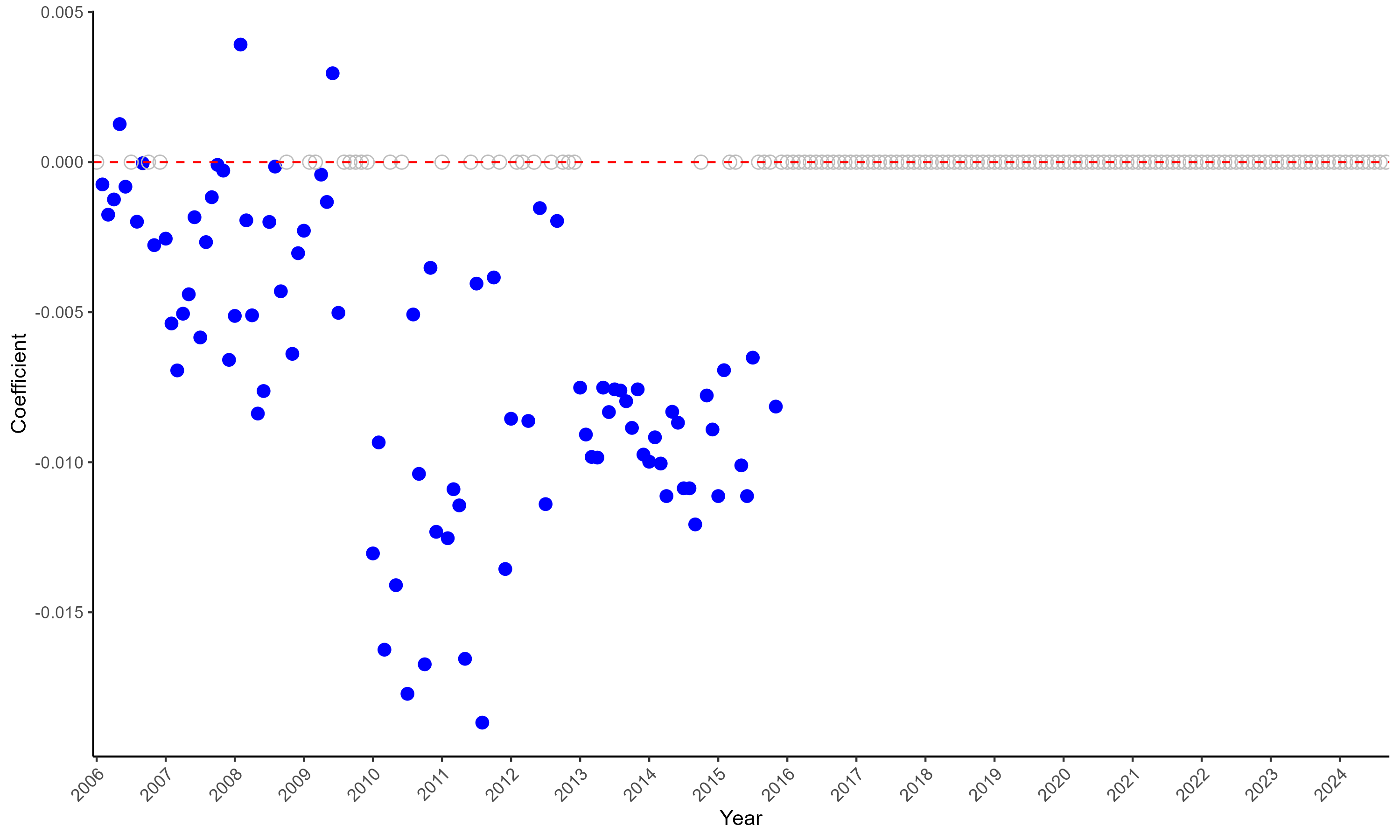}
        } \\
        \subfloat[COMPAPFFx]{
            \includegraphics[width=0.45\linewidth]{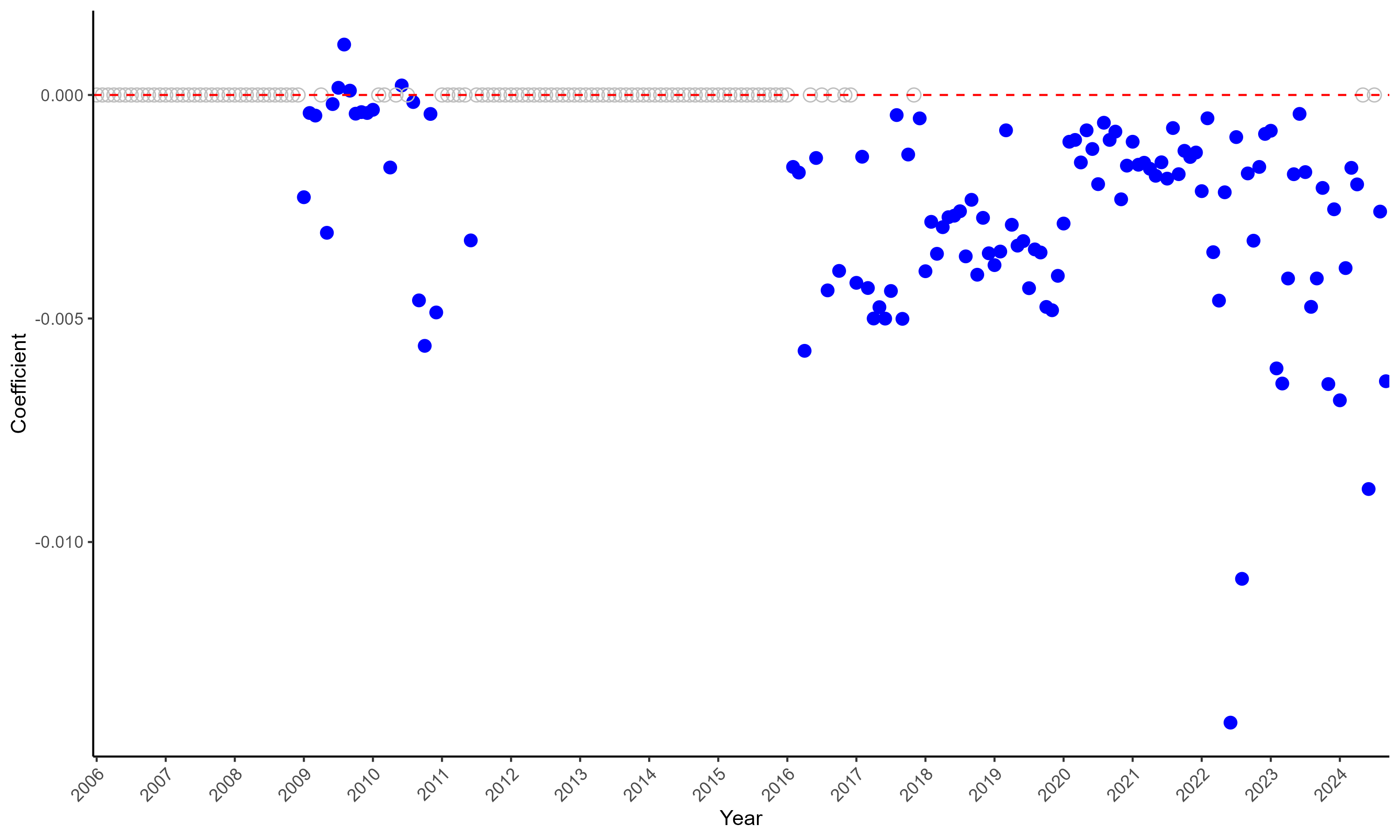}
        } & 
        \subfloat[TB6SMFFM]{
            \includegraphics[width=0.45\linewidth]{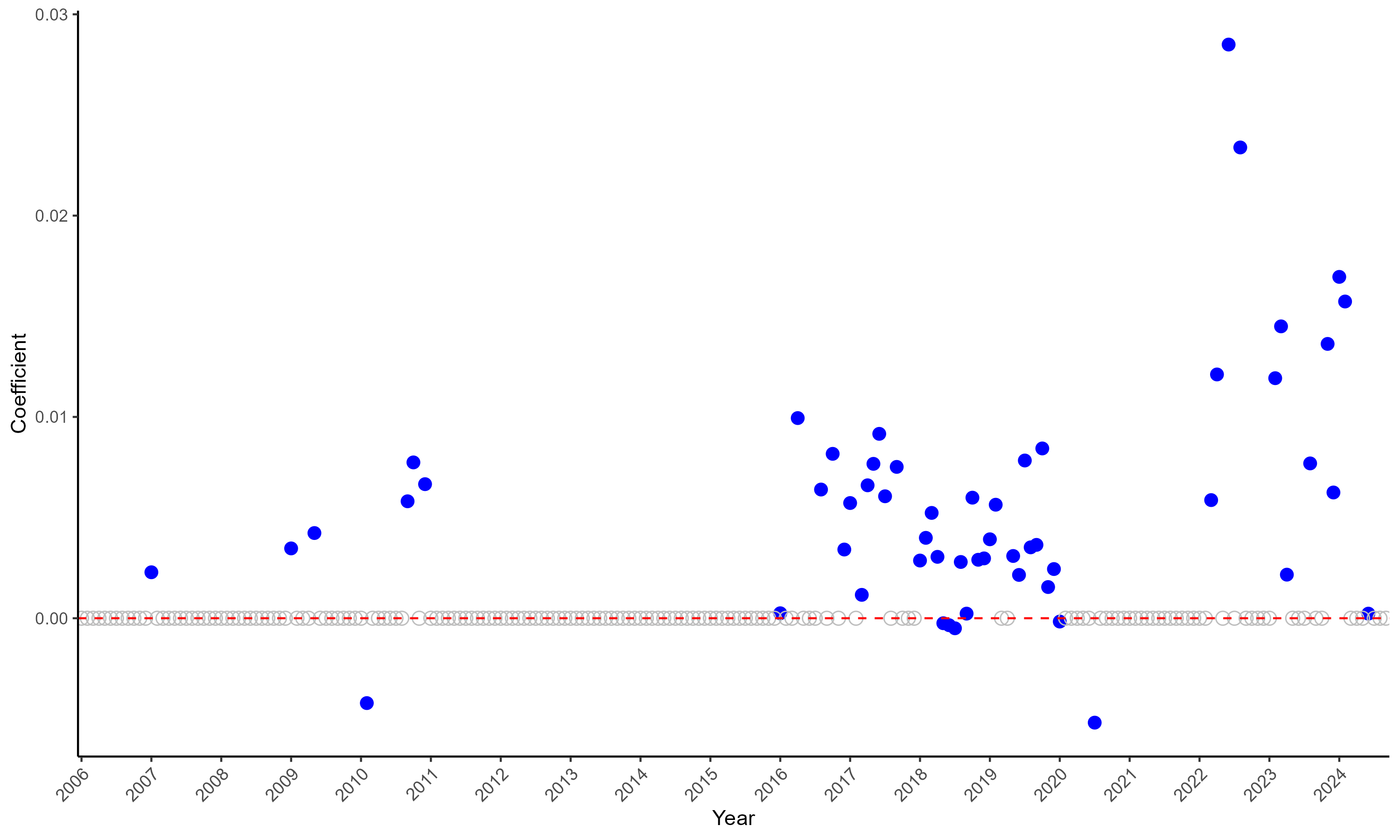}
        }\\
    \end{tabular}
    \caption{Coefficients of selected variables for $\tau = 0.05$ over time.}
    \label{Figure-QPCS-Coeffs-0.05}
    \footnotesize
    \begin{note*}
        \textit{Solid blue dots indicate periods of selection and show the corresponding value of the estimate. Hollow dots represent periods in which the variable was not selected. See Figure \ref{Figure-QPCS-Select-Heat-0.05} for definitions of the acronyms.}
    \end{note*}
\end{figure}

One of the benefits of QPCR over other ML methods is the interpretability of its results. Figure \ref{Figure-QPCS-Coeffs-0.05} shows coefficient estimates over time of some of the selected variables in the groups described above: CUMFNS (capacity utilisation), UNRATE (unemployment rate), CLAIMsx (initial claims), COMPAPFFx (three-month commercial paper minus the Federal Funds rate), HOUST (housing starts), and TB6SMFFM (six-month Treasury rate minus the Federal Funds rate). The first panel shows that the sign of the effect of capacity utilisation on growth vulnerability varies substantially over time, making it difficult to attach economic interpretation to this coefficient.

By contrast, the second and third panels show that increases in labour market slack—as measured by higher unemployment or initial claims—consistently worsen growth vulnerabilities. This is consistent with the view that supply shocks originating in the labour market increase recession risks. The fourth panel shows that that housing starts increased growth vulnerabilities, but only in the period surrounding the great financial crisis. Since approximately 2016, this predictor ceases to be informative about growth vulnerabilities, potentially reflecting improved macroprudential regulation. The fifth panel shows that increases in the commercial paper spread worsen growth vulnerabilities. This is consistent with the view that higher costs of firm financing--potentially reflecting heightened credit or liquidity risk--can translate into weaker investment and hence higher recession risk (see also \citet{gilchrist2012credit, gertler1999information}). The final panel shows that a widening spread between the six-month Treasury-bill rate and the Federal Funds rate is linked to lower growth vulnerabilities. This result is consistent with the view that yield-curve steepenings reflect expectations of economic recovery arising from anticipated monetary policy normalisation (see, e.g., \citep{estrella1998predicting}).

We hence conclude that the main predictors identified by QPCR are consistent with commonly held heuristics to assess recession risks. Importantly, however, the results further quantify these heuristics while controlling for the effect of all other predictors. For instance, we find that after controlling for all other variables, an increase in the unemployment rate of one standard deviation decreases the predicted lower-tail quantile ($\tau = 0.05$) of industrial production growth by up to half a percentage point. Similarly, after controlling for all other variables, we find that during the most recent hiking cycle, a one standard deviation increase in the commercial paper spread decreased the predicted lower-tail quantile ($\tau = 0.05$) of industrial production growth by one percentage point. Such quantification is particularly valuable for policymakers, as it clearly identifies which factors drive recession risk and measures their potential impact.

\pagebreak

\subsection{Growth at Risk decomposition and targeted indices \label{section-GaR-Decomp}}

An advantage of QPCR is that the predicted growth vulnerabilities can easily be decomposed into their underlying drivers. Letting $\mathcal G \subseteq\{1, \dots, p\}$ denote the indices corresponding to a particular set of variables (e.g., financial variables), we can write the total contribution of the predictors indexed in $\mathcal{G}$ to one-step-ahead growth vulnerabilities based on the sample $\{Y_{t+1}, X_{t}\}_{t = 1}^T$ and the conditioning predictor $X_{T+1}$ as
\[
\hat{Q}_{Y_{T+2}}^{QPCR, \mathcal G}(\tau|X_{\mathcal{G}, T + 1}) =\sum_{j\in \mathcal G}\hat{\beta}^{QPCR}_{j}X_{T+1, j}.
\]
We note that if $\mathcal{G}_1, \dots, \mathcal{G}_n$ form a partition of $\{1, \dots, p\}$, the quantile of $Y_{T+2}$ conditional on $X_{T+1}$, i.e., GaR, can be expressed as $\hat{Q}^{QPCR}_{Y_{T+2}}(\tau|X_{T + 1}) = \sum_{i = 1}^n\hat{Q}_{Y_{T+2}}^{QPCR, \mathcal G}(\tau|X_{\mathcal{G}_i, T + 1})$.

\begin{figure}[H]
    \centering
    \begin{tabular}{c}
        \subfloat{
    \includegraphics[width=0.9\linewidth]{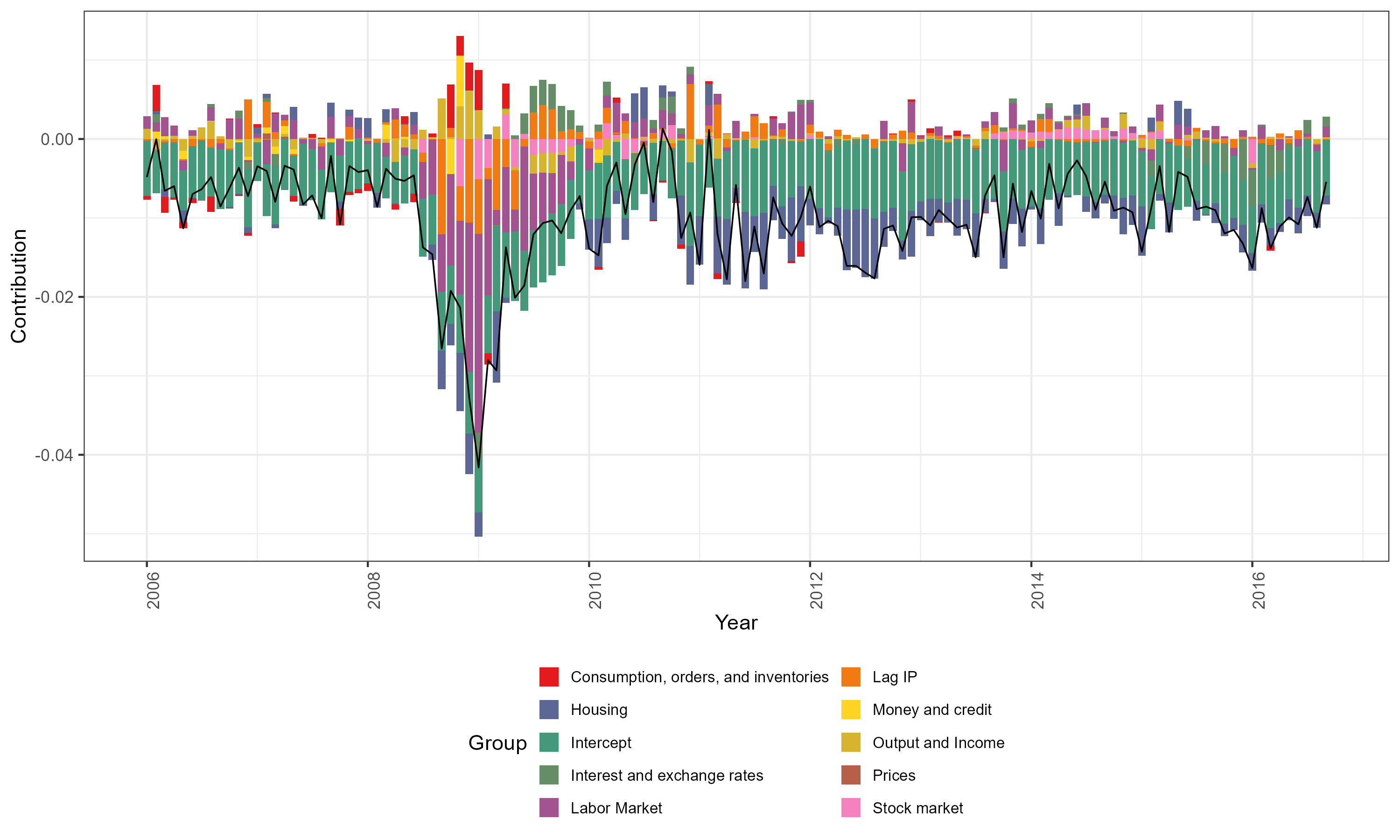}
        }\\
        \subfloat{
    \includegraphics[width=0.9\linewidth]{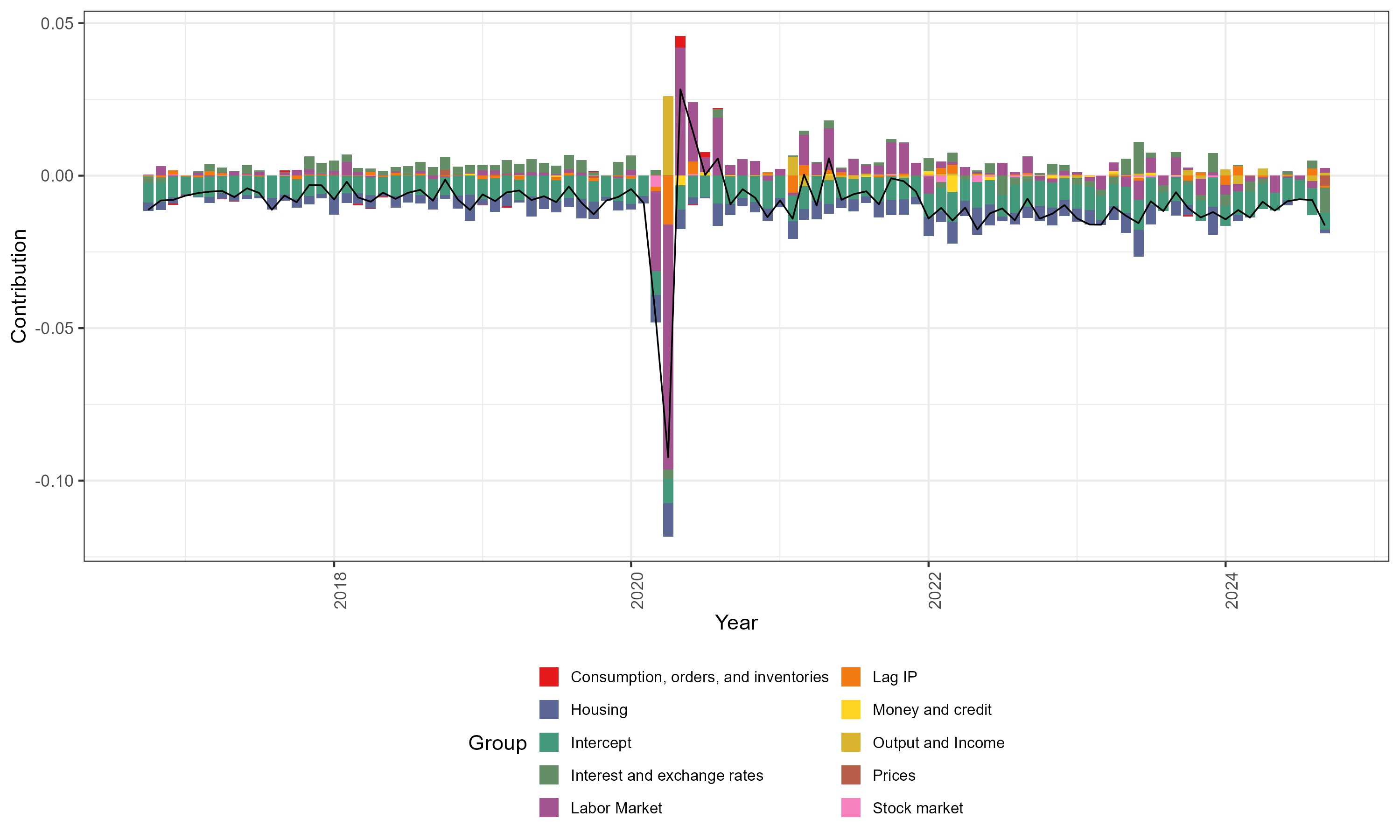}
        }\\
    \end{tabular}
    \caption{Decomposition of GaR into Fred-MD groups, $\tau = 0.05$.}
    \label{Figure-QPCS-Decomp-0.05}
    \footnotesize
    \begin{note*}
    \footnotesize
    \textit{Each stacked bar represents the contribution to the predicted conditional quantile of a specific group of indicators, $\hat{Q}_{Y_{T+2}}^{QPCR, \mathcal{G}_i}(\tau|X_{\mathcal{G}_i, T + 1})$, where $\mathcal{G}_i$ are the group labels in FRED-MD, as well as the lag of IP, and the constant. The black solid line shows the predicted conditional quantile, $\hat{Q}^{QPCR}_{Y_{T+2}}(\tau|X_{T+1}) = \sum_{i = 1}^n\hat{Q}_{Y_{T+2}}^{QPCR, \mathcal{G}_i}(\tau|X_{\mathcal{G}_i, T + 1})$.}
    \end{note*}
\end{figure}

Figure \ref{Figure-QPCS-Decomp-0.05} shows the GaR decomposition for $\tau = 0.05$, with contributions aggregated across Fred-MD group labels, as well as the lag of IP growth and a constant. During the GFC, the labour, housing, and stock markets were key sources of vulnerability. In the post-GFC period and leading up to the Covid-19 pandemic, vulnerabilities declined, with the constant term emerging as the main contributor to downside risk. During the Covid-19 pandemic, the labour market became the dominant driver of downside risk.

Decomposing GaR into the contributions from individual variables also makes it possible to create easily trackable and comparable summaries of our ML analyses in the form of sector-specific indices. We note that indices constructed in this way enjoy a series of favourable properties: they are targeted to particular quantiles of interest, they are predictive without suffering from look-ahead bias, and they isolate sector-specific effects without capturing unrelated information.

\begin{figure}[H]
    \includegraphics[width=1\linewidth]{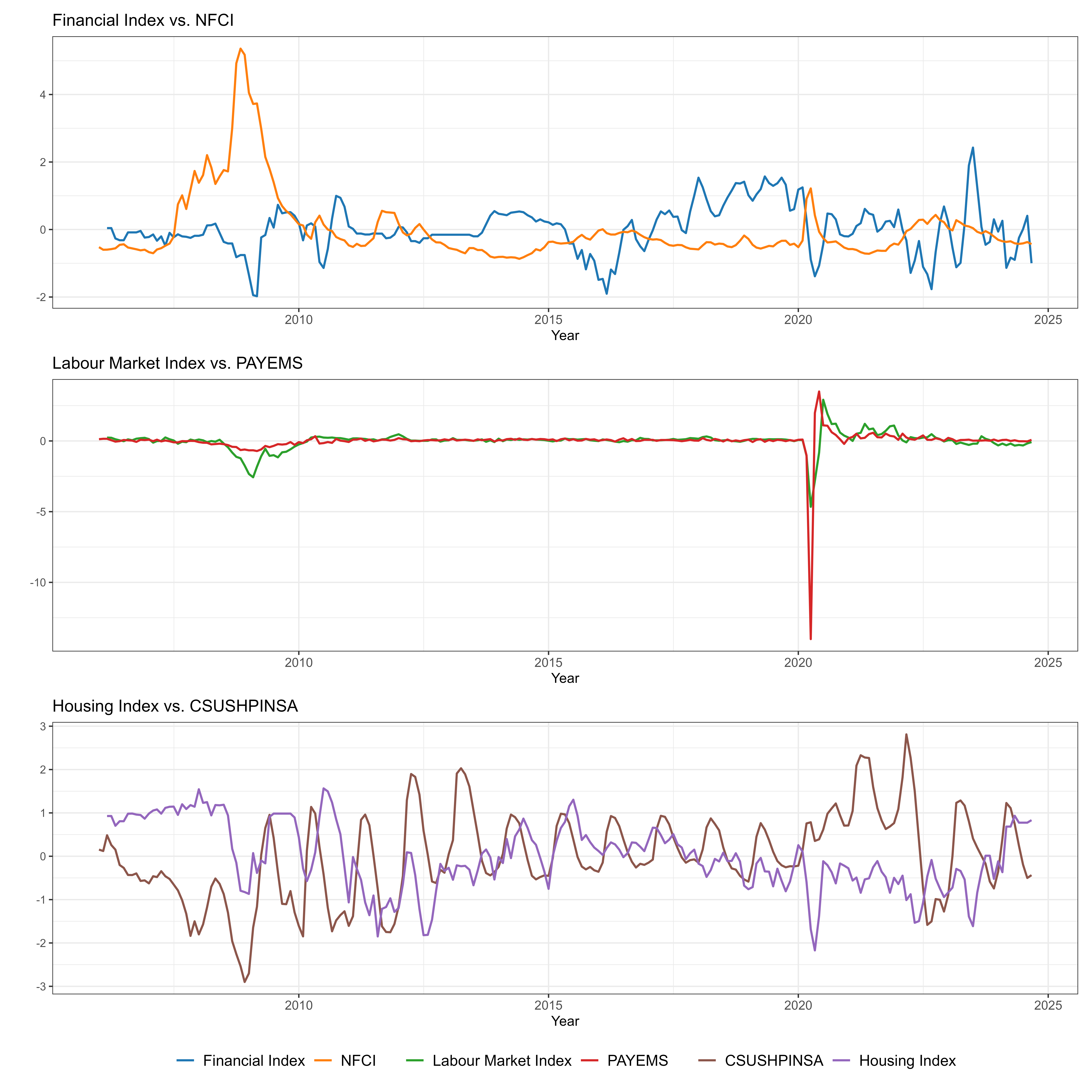}
    \caption{Targeted indices and commonly-used benchmarks.}
    \label{Figure-Targeted-Indices}
    \begin{note*}
    \footnotesize
    \textit{All indices are normalized to have zero mean and unit variance.}
\end{note*}
\end{figure}

For example, the first panel of Figure \ref{Figure-Targeted-Indices} shows the contribution of financial variables to the $\tau = 0.05$ quantile of one-period-ahead IP growth,\footnote{That is, $\hat{Q}_{Y_{T+2}}^{QPCR, \mathcal{G}_{financial}}(\tau|X_{\mathcal{G}_{financial}, T + 1})$, where $\mathcal{G}_{financial}$ contains the indices corresponding to the predictors: AAAFFM, BAA, BAAFFM, BOGMBASE, BUSLOANS, COMPAPFFx, CONSPI, CP3Mx, DTCOLNVHFNM, DTCTHFNM, FEDFUNDS, GS10, GS5, M2REAL, M2SL, NONBORRES, REALLN, S.P.500, S.P.PE.ratio, T10YFFM, T1YFFM, T5YFFM, TB3MS, TB3SMFFM, TB6SMFFM, TOTRESNS, and VIXCLSx.} averaged over three months for smoothness, together with the NFCI for comparison. Similarly to the NFCI, our targeted FCI worsens around the GFC, as well as the Covid-19 pandemic and the subsequent tightening cycle. However, in contrast to the NFCI, our targeted FCI improves between 2016 and 2020, suggesting that loose financial conditions helped decrease downside risks in this period. The second panel shows a similar plot for the contribution of labour-market variables,\footnote{The predictors included are: AWHMAN, AWOTMAN, CE16OV, CES0600000007, CES0600000008, CES1021000001, CES2000000008, CLAIMSx, CLF16OV, DMANEMP, HWIURATIO, MANEMP, NDMANEMP, PAYEMS, SRVPRD, UNRATE, USCONS, USFIRE, USGOOD, and USWTRADE.} together with nonfarm payrolls--a commonly used benchmark to assess real economic activity. Similarly to nonfarm payrolls, the targeted labour marked index worsens around the GFC and the Covid-19 pandemic. The third panel shows the targeted housing market index and the Case-Shiller house-price index--a commonly used benchmark to assess the US housing market.\footnote{The predictors included are: HOUST, HOUSTMW, HOUSTNE, HOUSTS, HOUSTW, PERMITMW, PERMITNE, PERMITS, and PERMITW.} The two series show some cyclical comovement, although there are protracted periods where they are asynchronous.

\begin{figure}[H]
{
    \includegraphics[width=0.78\linewidth]{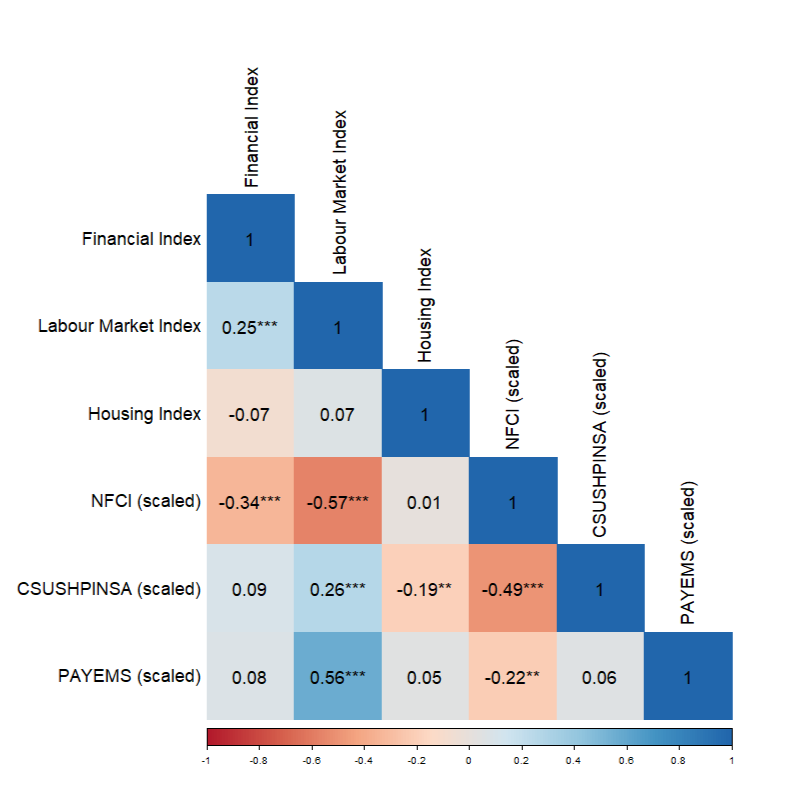}
    \caption{Pairwise correlations of indices and commonly-used benchmarks.}
    \label{Figure-Correlations-Indices}
    \begin{note*}
    \footnotesize
    \textit{The significance of the correlation between index or benchmark $i$ and $j$, $\text{r}_{i,j}$, is tested under the null hypothesis $H_0: \text{r}_{i,j} = 0$ using the statistic $z = \hat{r}_{i, j}\sqrt{(\tilde T - 2)/(1 - \hat{r}_{i, j}^2)}$, where $\hat{r}_{i, j}$ is the sample correlation between $i$ and $j$ and $\tilde T$ is the number of observations. *** $p$-value$< 0.01$, ** $p$-value$< 0.05$, * $p$-value$< 0.1$.}
    \end{note*}
    }
\end{figure}

Figure \ref{Figure-Correlations-Indices} shows the pairwise correlations of the targeted indices and benchmarks considered in Figure \ref{Figure-Targeted-Indices}. As suggested by the plots in Figure \ref{Figure-Targeted-Indices}, we find that our targeted FCI is most highly correlated with the NFCI, that our targeted labour market index is most highly correlated with payrolls, and that our targeted housing index is most highly correlated with the Case-Shiller house price index. We also find that the NFCI is most strongly correlated with our labour market index, and that it is significantly correlated with payrolls and the Case-Shiller house price index, suggesting that the NFCI is capturing non-financial information. Taken together, these findings highlight the value of constructing dedicated, sector‐specific indices using our proposed approach.

%% file: sections/conclusion_monthly.tex
\section{Conclusion}

We applied QPCR to predict downside risk to IP growth, and derived clear insights into growth vulnerabilities. Our analysis showed that financial, labour-market, and housing variables are the primary drivers of risks, with their relative importance evolving over time. Moreover, we validated and quantified commonly held views on key predictors of downside risks, including measures of slack, yield curve dynamics, and credit spreads. We also exploited the linear structure of QPCR to decompose downside risk into variable-by-variable contributions, which allowed us to construct sector-specific indices that are predictive of future vulnerabilities without suffering from look-ahead bias, while adjusting for the information of other sectors. Our framework can be readily adapted to other contexts, offering a flexible toolkit for GaR assessments.